\RequirePackage{lineno}
\documentclass[letterpaper,twocolumn,prl,aps,superscriptaddress,amsmath,amssymb,floatfix]{revtex4-2}
\usepackage{mathptmx}
\usepackage[latin9]{inputenc}
\usepackage{color}
\usepackage{amsmath}
\usepackage{amssymb}
\usepackage{graphicx}
\usepackage{esint}
	
\usepackage[unicode=true,
 bookmarks=true,bookmarksnumbered=false,bookmarksopen=false,
 breaklinks=false,pdfborder={0 0 1},backref=false,colorlinks=true]
 {hyperref}
\hypersetup{
 linkcolor=magenta,urlcolor=blue,citecolor=blue,pdfstartview={FitH},hyperfootnotes=false}

\makeatletter

\DeclareFontEncoding{LGR}{}{}
\DeclareRobustCommand{\greektext}{%
  \fontencoding{LGR}\selectfont\def\encodingdefault{LGR}}

\ProvideTextCommand{\~}{LGR}[1]{\char126#1}

\newcommand{\lyxmathsym}[1]{\ifmmode\begingroup\def\b@ld{bold}
  \text{\ifx\math@version\b@ld\bfseries\fi#1}\endgroup\else#1\fi}

\usepackage{textcomp}
\usepackage{epstopdf}

\usepackage{amsfonts}

\pdfpageheight\paperheight
\pdfpagewidth\paperwidth

\@ifundefined{textcolor}{}{%
 \definecolor{BLACK}{gray}{0}
 \definecolor{WHITE}{gray}{1}
 \definecolor{RED}{rgb}{1,0,0}
 \definecolor{GREEN}{rgb}{0,1,0}
 \definecolor{BLUE}{rgb}{0,0,1}
 \definecolor{CYAN}{cmyk}{1,0,0,0}
 \definecolor{MAGENTA}{cmyk}{0,1,0,0}
 \definecolor{YELLOW}{cmyk}{0,0,1,0}
}

\usepackage{xcolor}\usepackage{soul}
\definecolor{blue}{rgb}{0,0,1}
\definecolor{red}{rgb}{1,0,0}
\definecolor{green}{rgb}{0,1,0}

\makeatother

\begin{document}


\title{Non-reciprocal Cavity Polariton with Atoms Strongly Coupled to Optical Cavity}
\affiliation{State Key Laboratory of Quantum Optics and Quantum Optics Devices,and
Institute of Opto-Electronics, Shanxi University, Taiyuan 030006,
China}
\affiliation{CAS Key Laboratory of Quantum Information, University of Science and
Technology of China, Hefei, Anhui 230026, P. R. China}
\affiliation{Collaborative Innovation Center of Extreme Optics, Shanxi University,
Taiyuan 030006, China}
\affiliation{Institute of Big Data Science and Industry, Shanxi University, Taiyuan 030006, China}
\author{Pengfei~Yang}
\thanks{These authors contributed equally to this work.}
\affiliation{State Key Laboratory of Quantum Optics and Quantum Optics Devices,and
Institute of Opto-Electronics, Shanxi University, Taiyuan 030006,
China}
\affiliation{Collaborative Innovation Center of Extreme Optics, Shanxi University,
Taiyuan 030006, China}
\affiliation{Institute of Big Data Science and Industry, Shanxi University, Taiyuan 030006, China}
\author{Ming Li}
\thanks{These authors contributed equally to this work.}
\affiliation{CAS Key Laboratory of Quantum Information, University of Science and
Technology of China, Hefei, Anhui 230026, P. R. China}
\author{Xing Han}
\affiliation{State Key Laboratory of Quantum Optics and Quantum Optics Devices,and
Institute of Opto-Electronics, Shanxi University, Taiyuan 030006,
China}
\affiliation{Collaborative Innovation Center of Extreme Optics, Shanxi University,
Taiyuan 030006, China}
\author{Hai He}
\affiliation{State Key Laboratory of Quantum Optics and Quantum Optics Devices,and
Institute of Opto-Electronics, Shanxi University, Taiyuan 030006,
China}
\affiliation{Collaborative Innovation Center of Extreme Optics, Shanxi University,
Taiyuan 030006, China}
\author{Gang Li}
\email{gangli@sxu.edu.cn}

\affiliation{State Key Laboratory of Quantum Optics and Quantum Optics Devices,and
Institute of Opto-Electronics, Shanxi University, Taiyuan 030006,
China}
\affiliation{Collaborative Innovation Center of Extreme Optics, Shanxi University,
Taiyuan 030006, China}
\author{Chang-Ling~Zou}
\email{clzou321@ustc.edu.cn}

\affiliation{CAS Key Laboratory of Quantum Information, University of Science and
Technology of China, Hefei, Anhui 230026, P. R. China}
\affiliation{State Key Laboratory of Quantum Optics and Quantum Optics Devices,and
Institute of Opto-Electronics, Shanxi University, Taiyuan 030006,
China}
\author{Pengfei Zhang}
\affiliation{State Key Laboratory of Quantum Optics and Quantum Optics Devices,and
Institute of Opto-Electronics, Shanxi University, Taiyuan 030006,
China}
\affiliation{Collaborative Innovation Center of Extreme Optics, Shanxi University,
Taiyuan 030006, China}
\author{Yuhua Qian}
\affiliation{Institute of Big Data Science and Industry, Shanxi University, Taiyuan 030006,
China}
\affiliation{Key Laboratory of Computational Intelligence and
Chinese Information Processing of Ministry of Education, Shanxi University,
Taiyuan 030006, China}
\author{Tiancai Zhang}
\email{tczhang@sxu.edu.cn}

\affiliation{State Key Laboratory of Quantum Optics and Quantum Optics Devices,and
Institute of Opto-Electronics, Shanxi University, Taiyuan 030006,
China}
\affiliation{Collaborative Innovation Center of Extreme Optics, Shanxi University,
Taiyuan 030006, China}

\begin{abstract}
Breaking the time-reversal symmetry of light is of great importance
for fundamental physics and has attracted increasing interest in the study of non-reciprocal photonic devices. Here, we experimentally demonstrate a chiral cavity QED system with multiple atoms strongly coupled to a Fabry-P{\'e}rot cavity. By polarizing the internal quantum state of the atoms, the time-reversal symmetry of the atom-cavity interaction is broken. The strongly coupled atom-cavity system can be described by non-reciprocal quasiparticles, i.e., the cavity polariton. When it works in the linear regime, the inherent nonreciprocity makes the system work as a single-photon-level optical isolator. Benefiting from the collective enhancement of multiple atoms, an isolation ratio exceeding 30~dB on the single-quanta level ($\boldsymbol{\sim0.1}$ photon on average) is achieved. The validity of the non-reciprocal device under zero magnetic field and the reconfigurability of the isolation direction are also experimentally demonstrated. Moreover, when the cavity polariton works in the nonlinear regime, the quantum interference between polaritons with weak anharmonicity induces non-reciprocal nonclassical statistics of cavity transmission from coherent probe light.
\end{abstract}

\maketitle

\section{Introduction}
Magnet-free optical non-reciprocal devices, in which the light propagates 
non-reciprocally from opposite directions, have great application potential in photonic information processing~\cite{Jalas2013,Asadchy2020}. They also allow the realization of artificial gauge fields for photons and the simulation of interesting effects that were previously only available for electrons~\cite{Goldman2014,Hey2018}. The optical nonreciprocity (ONR) intrinsically relies on the time-reversal (T) symmetry breaking of photon propagation, which was traditionally realized via a notable magneto-optical effect with the precondition of an intense external bias DC-magnetic field~\cite{Potton2004}. However, the strong magnetic field and the associated nonreconfigurable and nonswitchable functionality greatly limits the applications~\cite{Bi2011}. Therefore, many efforts have been devoted to magnet-free ONR, including spatiotemporal modulation of dielectric permittivity~\cite{Yu2009,Wang2013, Horsley2013,Sounas2017,Ramezani2018}, synthetic magnetic field~\cite{Tzuang2014,Yuan2015,Fang2017}, optics frequency conversion processes~\cite{Kamal2010,Estep2014,Hua2016,Song2021,Li2023,Chang2014,Xia2018,Guo2016}, optomechanics~\cite{Shen2016,Ruesink2016,Verhagen2017,Kim2015,Kang2011,Kittlaus2021}, chiral light-matter interaction~\cite{Shomroni2014,Sollner2015,Bechler2018}, the Doppler effect~\cite{Zhang2018,Lin2019,Liang2020, Dong2021} and light-induced magnetization~\cite{Hu2021} in atomic media, coherent interference of spin-waves~\cite{Lu2021}, spinning resonators~\cite{Maayani2018}, etc. One of the most adopted approaches is to utilize coherent nonlinear optical effects, by which the direction-dependent transmission of the probe light is realized with a coherent external drive in a fixed direction~\cite{Kamal2010,Estep2014,Hua2016,Song2021,Li2023,Chang2014,Xia2018,Guo2016} or a refractive index modulation with an effective momentum ($\boldsymbol{p}$)~\cite{Yu2009,Sounas2017,Ramezani2018,Wang2013,Horsley2013}.
The T-symmetry is broken as $\mathcal{T}\boldsymbol{p}\mathcal{T}^{-1}=-\boldsymbol{p}$
with $\mathcal{T}$ being the time-reversal operator. All-optical nonreciprocity 
with such a mechanism has been experimentally
verified in optomechanics and nonlinear microresonator platforms~\cite{Shen2016,Guo2016,Hua2016,Sohn2018},
in which strong bias AC-driving fields are needed.

Alternatively, since $\mathcal{T}\boldsymbol{S}\mathcal{T}^{-1}=-\boldsymbol{S}$
for the spin operator $\boldsymbol{S}$, the T-symmetry is naturally
broken by preparing the internal spin state of atoms or emitters with $\boldsymbol{S}\neq-\boldsymbol{S}$.  
Therefore, when the photon interacts with atoms, assisted by the selection rules, the non-reciprocal transmission
of the photon propagating along the $z$-direction could be realized by preparing the atoms to a ground state with biased spin state $S_{z}$. Recently, the
strong coupling between single atoms or single quantum dots and chiral photons has been studied experimentally
in a whispering-gallery-mode (WGM) microresonator~\cite{Junge2013} and nanowaveguide~\cite{Sayrin2015},
where the non-reciprocal transmission 
of light fields can be realized via the unequal interaction strength between the emitters and photons with different chirality. The isolation~\cite{Sollner2015} and circulation~\cite{Scheucher2016} of single photons have been demonstrated. In these experiments, the intrinsic chiral property of the evanescent field of the WGM microresonator or the nanowaveguide is adopted, but the achievable isolation ratio is limited to 13\,dB due to the small atom number.


In this paper, the chiral light-matter interaction, where the interaction between atoms and two circular photons with orthogonal polarization are different, is introduced into the conventional cavity QED with an optical Fabry-P{\'e}rot cavity (FPC) experimentally. The collective interaction between multiple 
atoms with a miniature high-finesse optical FPC greatly boosts the cooperativity of the system, and the strongly coupled photons and atoms constitutes the hybrid quasiparticles, which are called ``cavity polaritons". The non-reciprocal polariton is then realized by preparing the atoms in a spin-polarized internal state with total spin $S \neq 0$, which asymmetrically couple to the two orthogonal circularly polarized cavity modes. 
The polariton inherently breaks the T-reversal symmetry 
and gives rise to non-reciprocal vacuum Rabi splitting spectra. By introducing polarizers and waveplates outside the cavity, the circular polarization of inputs to the cavity is locked with the direction of the inputs, and thus, the non-reciprocal polaritons enable an optical isolator with an isolation over $30\,\mathrm{dB}$ and a bandwidth exceeding $10\,\mathrm{MHz}$. Compared to our previous demonstration of ONR by using few-atom nonlinear bistability~\cite{Yang2019}, which works on the few-photon level, this work is based on a vacuum-induced quantum process and works on the single-quanta level ($\boldsymbol{\sim0.1}$ intracavity photon number on average). The isolation can be further enhanced by increasing the number of atoms. The direction of the isolation is reconfigurable by switching the internal state of spin-polarized atoms.
The device is capable of working under a zero magnetic field with the aid of a circularly polarized optical pumping field propagating along the cavity to maintain the polarization of the atom.

Additionally, by employing the weak anharmonicity of the polaritons, nonclassical features of light can be generated non-reciprocally from coherent input light because of the quantum interference between polaritons. In particular, sub-Poissonian versus super-Poissonian photon statistics, or bunching versus antibunching effects, are observed when probing the system from different directions.
Taking advantage of the few-atom cavity QED platform, the adjustable nonlinearity of non-reciprocal polaritons allows studies of potential non-reciprocal quantum effects and nonlinear dynamics. This new quasiparticle holds great potential for exploring quantum nonreciprocity in photonics~\cite{Xia2014,Huang2018} and quantum network applications~\cite{Borregaard2019,Lodahl2017}, chiral photophysics of molecules~\cite{Sun2022} and new topological effects of polaritons~\cite{Lodahl2017,Houck2012,Angelakis2017,Zhao2019,Lyons2021}. 


\section{Principle of non-reciprocal cavity polaritons}

\begin{figure}
\centering{}\includegraphics[width=1\columnwidth]{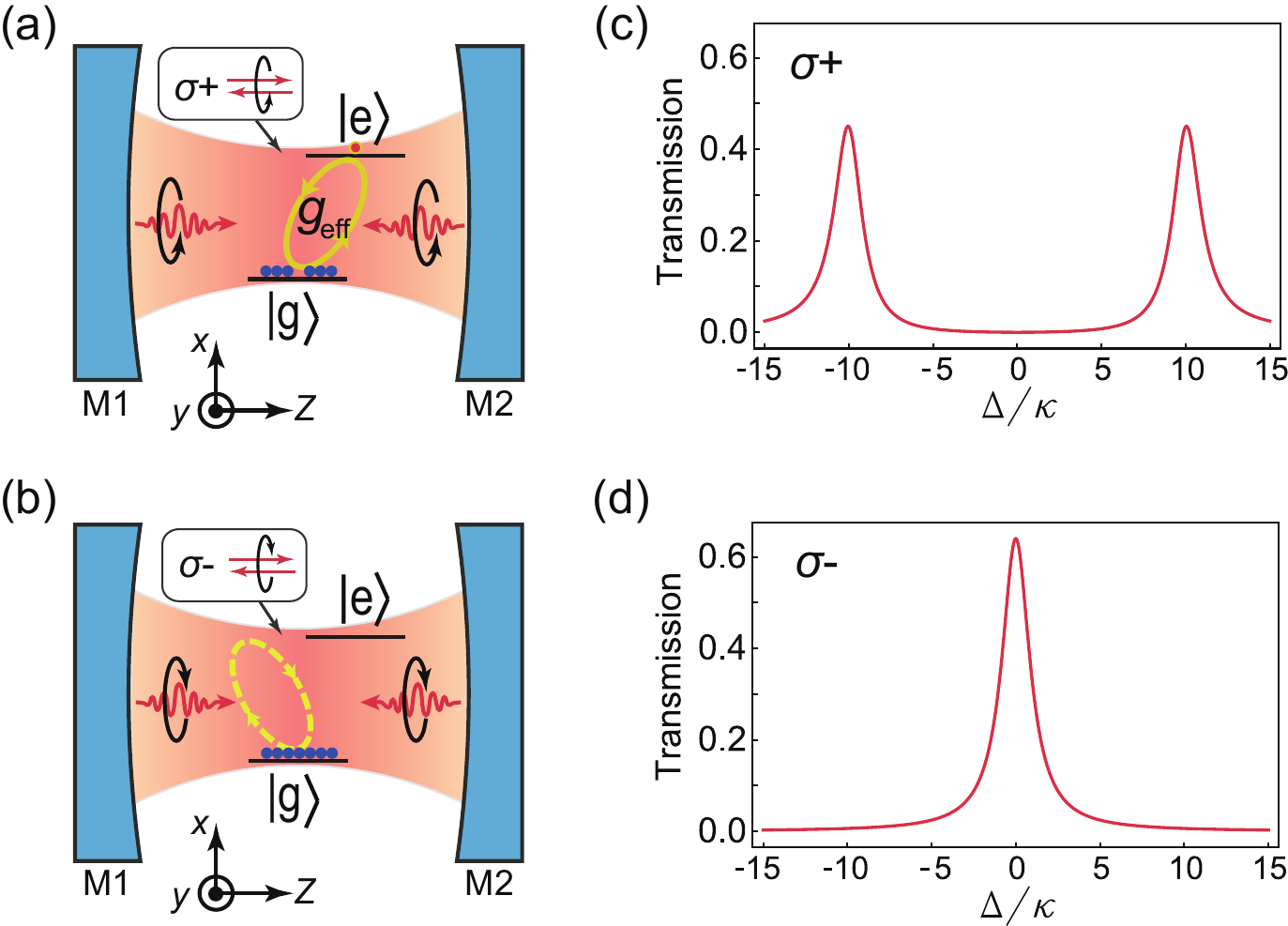} \caption{\label{fig1} Chiral interaction between atoms and a Fabry-Perot cavity.
a) and b) are schematic illustrations of the chiral
interaction, in which the $\sigma_{+}$-mode of the cavity is strongly coupled to the atoms
(a) but the $\sigma_{-}$-mode is transparent (b) due to the absence of the atomic transition. Here, the $\sigma_+$-($\sigma_-$-) polarized field is defined by the rotating direction of the electric field with respect to the direction of the quantization axis ($z$-axis), which could be either an LCP field or an RCP field defined in optics (see Section S1 in the Supporting Information for details). In a) and b), $(+z)$ is taken as the quantization axis along the cavity. The propagation direction of the signal light is not presented because the interaction with the atom has no connection with the propagation direction.  c) and d) are the theoretical
predictions of the corresponding transmission spectra for the non-reciprocal interacting system by weak signals with polarizations of $\sigma_{+}$ (c) and
$\sigma_{-}$ (d), respectively. In c), two polariton states (the hybrid atom-photon states) are obviously observed. The parameters of $C_{+}={50}$, $C_{-}=0$, $\kappa=\gamma_{+}=1$, and $g_{\text{eff}}/\kappa=10$ are used for the calculation.}
\end{figure}

Figure~\ref{fig1} schematically illustrates the concept of 
the chiral atom-cavity system. A cavity QED setup
with neutral atoms coupled to an FPC~\cite{Raimond2001,Reiserer2015,Yang2019} is adopted. The cavity supports two degenerate circularly polarized optical modes ($\sigma_+$ and $\sigma_-$), which asymmetrically couple to the atoms. Here, the $\sigma_+$- and $\sigma_-$-polarized fields are defined by the rotating direction of the electric field with respect to the direction of the quantization axis ($z$-axis). It is worth noting that the $\sigma_{\pm}$ polarization is not related to the propagation direction of the photon, and the $\sigma_{\pm}$-polarized field could be either a left-handed circularly polarized (LCP) field or a right-handed circularly polarized (RCP) field defined in optics, and the $\sigma_{+}$-polarized and $\sigma_{-}$-polarized photons are chiral to each other (see Section S1 in the Supporting Information for details). In the following, we refer to the $\sigma_{\pm}$ notation for describing the chiral photon-atom interaction since the corresponding atomic transitions with respect to the quantization axis couple with the $\sigma_{\pm}$-polarized photon according to the selection rules.

As shown in Figure~\ref{fig1}a and b, the $\sigma_+$-polarized light field couples to the atomic transition $|g\rangle \leftrightarrow |e\rangle$, whereas no atomic transition is available for the $\sigma_-$-polarized light field. This situation can be easily found in an atom by preparing the atom to a ground spin-polarized state with Zeeman quantum number $m_{F}=-F$ (with the hyperfine quantum number $F\neq0$) and using the atomic transition $|F\rangle \leftrightarrow |F'\rangle$ (with the hyperfine quantum number $F' \le F$).
The propagation direction of the signal light is not presented in Figure~\ref{fig1} because the interaction with the atom has no connection with the propagation direction. 
The collective atom-cavity coupling cooperativity for the $\sigma_+$-polarized mode is much greater than unity, and cavity polaritons between the photons and atoms are then constituted. 
The cavity polaritons, shown as quasiparticles with annihilation operator $p_{u,l}=\frac{1}{\sqrt{2}}\left(a_{+}\pm b_{+}\right)$
at the low excitation limit ($u$ and $l$ denote the upper and lower polaritons respectively; see Section S5
in the Supporting Information for details), have eigenfrequencies detuned from the
bare cavity by $\pm g_{\mathrm{eff}}$. Here, $a_{+}$ and $b_{+}$
denote the bosonic operator of the $\sigma_{+}$-polarized cavity mode
and the corresponding collective pseudospin operator for atomic transition,
respectively, and $g_{\mathrm{eff}}$ is the effective coupling strength.
Hence, the system shows two polariton states in the spectra (two peaks in Figure~\ref{fig1}c),
and both emit only $\sigma_{+}$-polarized photons. When the time is reversed, the $\sigma_{+}$-polarized photon becomes $\sigma_{-}$. The
atoms are transparent to the $\sigma_{-}$-polarized photons due to
the absence of atomic transitions (Figure~\ref{fig1}b. Thus, the
mode remains as a bare cavity mode (Figure~\ref{fig1}d). Comparing 
these two cases, it is obvious that the T-symmetry is broken for the atom-cavity interacting system.


The T-symmetry breaking of the system would break the Lorentz reciprocity of forward ($+z$) and backward ($-z$) propagating probe light with $\sigma_{+}$ 
and $\sigma_{-}$ polarizations, respectively. Under
the linear approximation (Section S4 in the Supporting Information), the transmission
of the whole system reads 
\begin{equation}
T_{\pm}=\frac{4\kappa_{1}\kappa_{2}}{\kappa^{2}}\left|\frac{1}{i\Delta/\kappa+1+2C_{\pm}/(i\left(\Delta+\Delta_{\mathrm{ac}}\right)/\gamma+1)}\right|^{2},\label{eq1}
\end{equation}
where $\pm$ indicates the probe light condition $\left(+z,\sigma_{+}\right)$
or $\left(-z,\sigma_{-}\right)$. Here, $\kappa$ and $\kappa_{1(2)}$
denote the total decay rate of the cavity mode and the external coupling
rate to the cavity through mirror M1 (M2), respectively. $\Delta$ ($\Delta_{\mathrm{ac}}$)
is the frequency detuning of the cavity modes to probe laser
(atomic transition), and $C_{\pm}=g_{\mathrm{eff},\pm}^{2}/\left( 2\kappa\gamma \right)$
is the parameter of cooperativity for two $\sigma$-polarized cavity modes with $g_{\mathrm{eff},\pm}=g_{\pm}\sqrt{N_{\text{eff}}}$,
where $N_{\text{eff}}$ is the effective intracavity atom number
and $g_{\pm}$ is the coupling strength of the $\sigma_{\pm}$ cavity mode with one atom and depends on the atomic population distribution on the ground states.
Beyond the linear approximation, the cavity polaritons actually exhibit anharmonicity for a finite atom number,
with an effective Kerr coefficient $\mp g_{\mathrm{eff}}/2N_{\mathrm{eff}}$
for $p_{u,l}$ (Section S5
in the Supporting Information). In general, the
atoms in thermal equilibrium would give a uniform population on ground state Zeeman levels, and the spin polarization would be absent with $S=0$. Both $\sigma_{+}$- and $\sigma_{-}$-polarized light couple to atoms with the same strengths with $C_{+}=C_{-}$; therefore, the system is reciprocal. For a polarized spin
state with a slanted population distribution on Zeeman levels (nonuniform distribution with $S\neq 0$),
Eq. (\ref{eq1}) predicts distinguishable spectra as $C_{+}\neq C_{-}$,
manifesting the nonreciprocity of the system.

\section{Results} 
\subsection{Experimental setup}

\begin{figure*}
\centering{}\includegraphics[width=0.7 \textwidth]{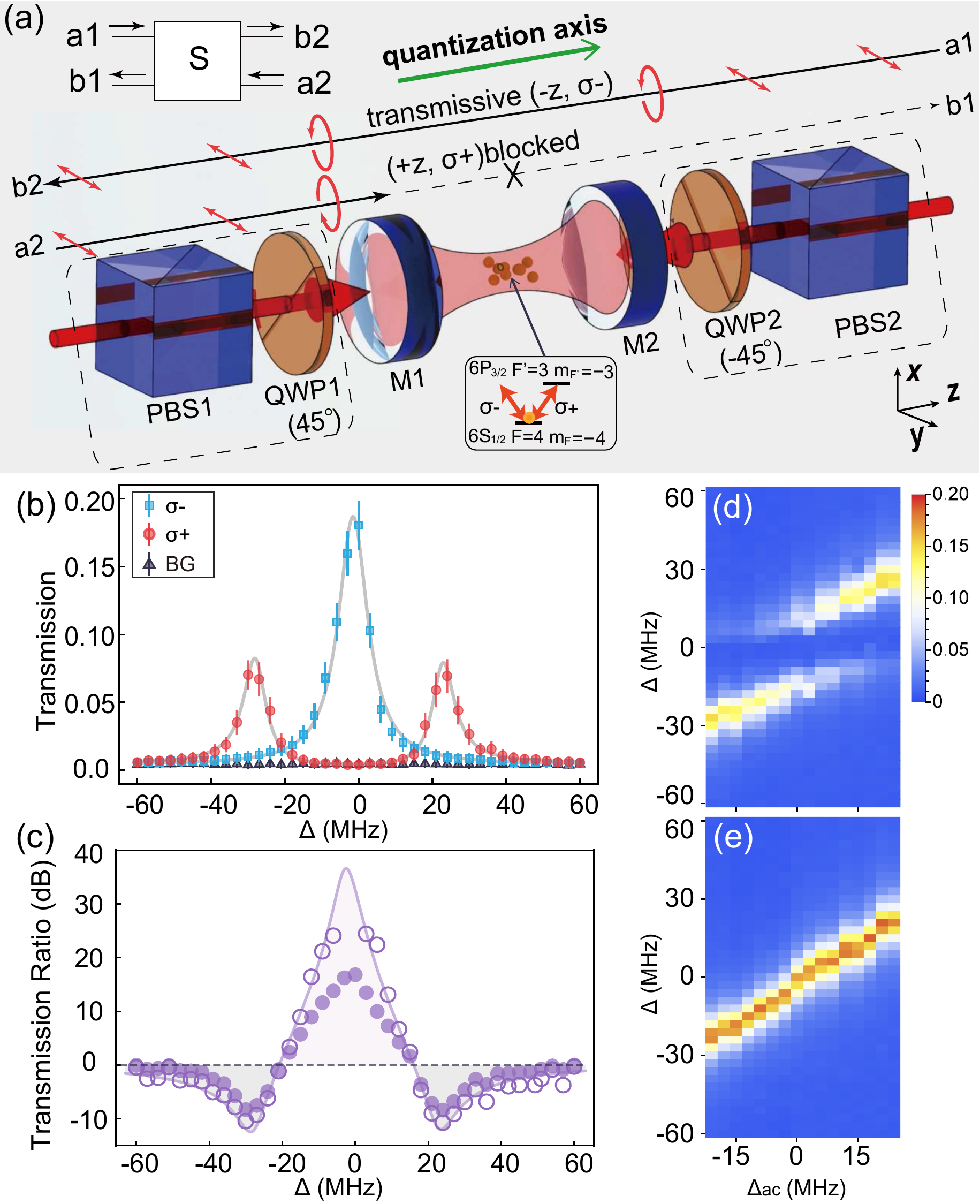} \caption{\label{fig2} Experimental demonstration of the chiral cavity QED system and non-reciprocal polaritons. a) A sketch of the experimental setup. A strongly coupled cavity QED system between multiple maximally spin-polarized Cs atoms and the $\sigma_+$-polarized cavity mode is placed between two sets of PBSs and
QWPs, where the locking of optical momentum-spin and the propagation direction can be realized. The quantization axis is along the cavity [green arrow, also the $(+z)$ direction]. Red arrows indicate local optical polarization. A $\sigma_{+}$ polarized light propagating along the $(+z)$ direction is blocked, while a $\sigma_{-}$ polarized light propagating along the $(-z)$ direction is transmitted. Insets: diagram of the atom energy levels (bottom) and scattering S matrix of the system (upper left corner). b) The measured transmission spectra with $\sigma_{+}$ (red circles)
and $\sigma_{-}$ (blue squares) polarized probes propagating along opposite directions. 
The background noise (black triangles) is
shown as a comparison. The error bars correspond to one standard deviation of multiple measurements. The solid curves are the theoretical fittings by Eq.~\ref{eq1}. 
c) The measured transmission ratio between the $\sigma_{\mp}$-polarized probe versus detuning. The parameters are the same as in b). 
The open and solid circles are the results with and without background correction, respectively, and the purple curve is the theoretical result. 
d) and e) Two-dimensional spectra of the system
probed by $\sigma_{+}$ (d) and $\sigma_{-}$ (e) polarized light with the probe-cavity detuning $\Delta$ scanned
at different cavity-atom detunings $\Delta_{\text{ac}}$.
}
\end{figure*}

The experiments are performed on a cavity QED system with multiple cesium (Cs) atoms coupled to a miniature high-finesse FPC~\cite{Yang2019}. A sketch of the experimental system is shown in Figure~\ref{fig2}a. A strongly coupled cavity QED system with multiple maximally spin-polarized Cs atoms is placed between the optical momentum-spin locking apparatuses, which consist of two sets of polarization beam splitters (PBSs) and quarter wave plates (QWPs), as shown in the dashed boxes of Figure~\ref{fig2}a. 
The axis of the QWP is oriented $45^\circ$ to the polarization of the transmitting light field from the PBS, and the optical axes of the two QWPs 
are orthogonal to each other. The probe light field with horizontally linear polarization transmits the apparatuses in both directions (from a1 to b2 and from a2 to b1), and the polarization is locked to the propagating direction in the region between two QWPs, i.e., the probe light can only be in either $\left(+z,\sigma_{+}\right)$ or $\left(-z,\sigma_{-}\right)$ states with a fixed quantization axis (green arrow). Without special declarations, a weak 3-Gauss magnetic field is used to define the quantization axis. 
Without the cavity QED system or if a cavity QED system with atomic spin $S=0$ is considered, the optical system is reciprocal. However, the reciprocity will be broken if a non-reciprocal cavity QED system (with atomic spin $S\neq 0$) is placed between the two QWPs.
The whole system resembles a commercial isolator except that the Faraday rotator is replaced by a non-reciprocal cavity QED system. 
It is worth noting that, like the Faraday-rotator-based optical isolator, this device can also be used as a four-port circulator, and here, we only use it as an isolator by neglecting the reflections from the two PBSs. 

The FPC is assembled 
by two concave mirrors with curvature radii of 100 mm, 
and the cavity length is 335 $\mu$m.
The concave surfaces are highly reflective, and the cavity has a finesse
of $6.1 \times 10^4$. A 1064 nm optical dipole trap (ODT) laser beam (horizontal) with a waist of 36 $\mu$m is used to load cold atoms from the MOT and transfer the atoms to the cavity. To prepare atoms to the maximally
spin-polarized internal state, $S_{z}=-4$ ($|6S_{1/2},F=4,m_{F}=-4\rangle$), a $\sigma_{-}$-polarized 459 nm pump laser with a beam waist of 550 $\mathrm\mu$m along the cavity axis and a linearly polarized 894 nm repump laser beam perpendicular to the cavity axis are used. The 459-nm and 894-nm lasers are resonant to Cs transitions $|6S_{1/2},F=4\rangle\leftrightarrow|7P_{1/2},F'=4\rangle$ and $|6S_{1/2},F=3\rangle\leftrightarrow|6P_{1/2},F'=4\rangle$, respectively. The coupling strength between the $\sigma_{+}$-mode of the cavity and single atom is $g_{+}=2\pi\times1.7$ MHz, and the decay rates of the cavity and the atom are $(\kappa,\gamma_{+})=2\pi\times(3.7,2.6)$
MHz. When multiple atoms couple to the cavity, the collective coupling strength 
would surpass the decay rates of the cavity and atom and make the system work in the strong-coupling regime. 

\subsection{Demonstration of non-reciprocal polaritons} 

To demonstrate the non-reciprocal polariton, 
the bare cavity mode is tuned to be resonant to the atomic transition $|6S_{1/2},F=4,m_{F}=-4\rangle\leftrightarrow|6P_{3/2},F'=3,m_{F'}=-3\rangle$.
We thus realize the ideal model shown in Figure~\ref{fig1}a and b as only the $\sigma_{+}$-mode couples to the atoms owing to the absence of an excited state for the $\sigma_{-}$-transition ($C_{-}=0$) (the energy levels can be found in the inset of Figure~\ref{fig2}a). The scattering matrix ($\mathbf{S_{I}}$) for the system is $\left(\begin{matrix}0&0\\\frac{4\kappa_1\kappa_2}{\kappa^2}&0\\\end{matrix}\right)$ with $C_{+}\gg 1$ and $\Delta=\Delta_{\text{ac}}=0$ (see Section S4 in the Supporting Information for details). 

By probing the system with $\sigma_{\pm}$-polarized light, we obtain the non-reciprocal vacuum Rabi splitting spectra, as shown in Figure~\ref{fig2}b, which agree well with the theoretical predictions from Eq.~(\ref{eq1}) (Figure~\ref{fig1}c and d). Vacuum Rabi splitting is observed only from the $\sigma_{+}$-polarized probe, which indicates collective cooperativity $C_{+}=33.8(0.2)$ and an effective atom number $N_{\text{eff}}=230.0(1.7)$. The atoms are optically pumped to the target spin state $|6S_{1/2},F=4,m_{F}=-4\rangle$ with a fidelity of approximately 95\%; thus, $C_{-}\ll C_{+}$ is still attained. The non-reciprocal polariton is further verified by the two-dimensional spectra (Figure~\ref{fig2}d and e), where the characteristic avoid-crossing spectra for polaritons can be accessed with only $\sigma_{+}$-polarized light as atom-cavity detuning $\Delta_{\text{ac}}$ being scanned.

\begin{figure*}
\centering{}\includegraphics[width=0.8\textwidth]{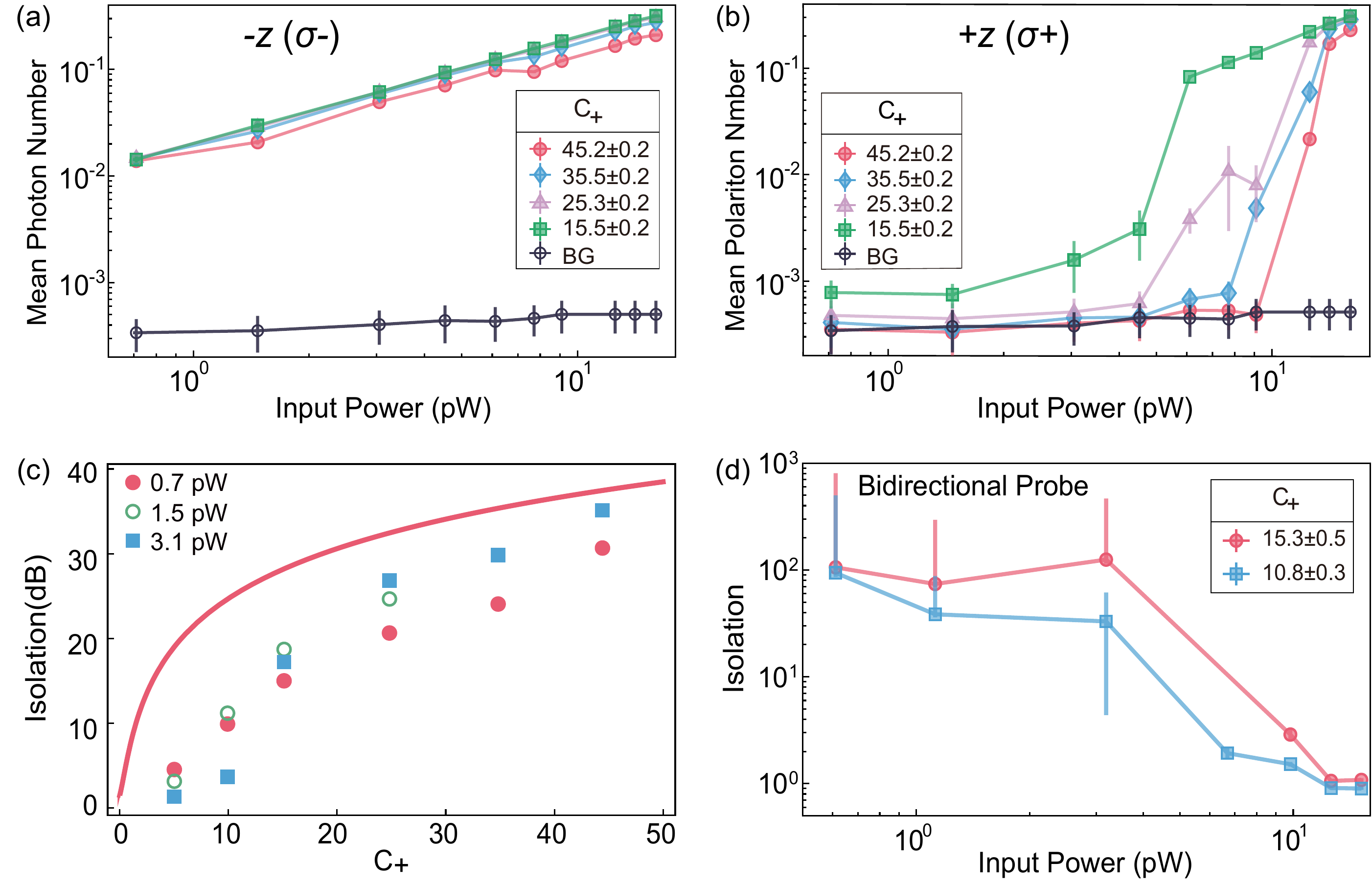} \caption{\label{fig3} Optical isolation based on the non-reciprocal cavity polaritons. a) and b) The measured intracavity mean photon number under a series of cooperativity $C_{+}$ versus input power
of the probe propagating along the $(-z)$ and $(+z)$ directions, respectively. c) The measured
isolation versus cooperativity $C_{+}$ under a series of input powers. The red solid circles, green open circles, and blue solid squares are for 0.7 pW , 1.5 pW, and 3.1 pW, respectively. d) The performance of the isolator with the forward and backward lights existing simultaneously, where the two lights have the  same power. The corresponding cooperativities $C_{+}$ are $15.3(0.5)$ (red circles) and $10.8(0.3)$ (blue squares). The error bars correspond to one standard deviation.}
\end{figure*}

\begin{figure}
\centering{}\includegraphics[width=1 \columnwidth]{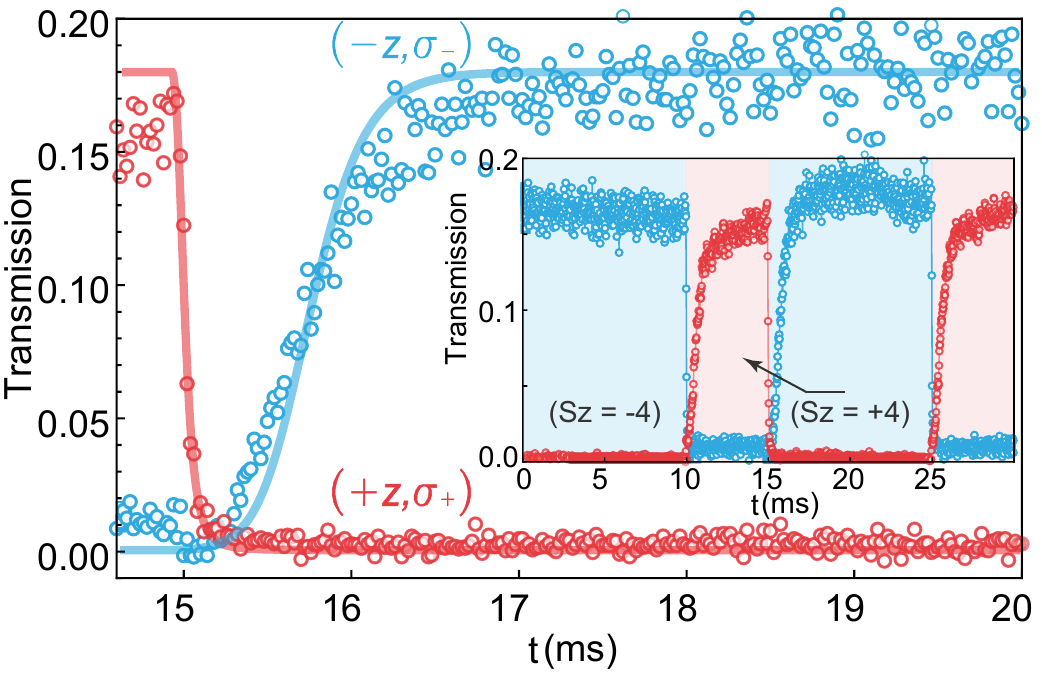} \caption{\label{fig4} Reconfiguration of the isolation direction. Measured transmission as a function of time accompanied by atomic state switching between $S_{z}=-4$ and $+4$, where $C_\text{+(-)}=20.5$. The blue open circles (experiment) and solid line (theory) are for forward light $(-z, \sigma_-)$, and the red ones are for backward light $(+z, \sigma_+)$. The switching times are $705\pm19$ and $43\pm2$ $\mu$s for the shut-off and turn-on processes, respectively. The inset shows the switching of the isolator continuously. The experimental data points are the average of 230 trials with 0.01 mean probe photons inside the FPC.}
\end{figure}

In Figure~\ref{fig2}b, if we focus on the frequency region around the atomic (or cavity)
resonance, the transmission of the $\sigma_{+}$-polarized probe is blocked, but the $\sigma_{-}$-polarized probe 
is transmitted. Since the two orthogonally polarized probe lights could only be injected into the cavity QED system from
opposite directions, the whole system actually operates as an optical isolator. 
The isolation is defined by the transmission
ratio $\mathcal{I}=T_{-}/T_{+}$. Figure~\ref{fig2}c presents the
measured $\mathcal{I}$ versus the probe detuning ($\Delta$) with
$\Delta_{\mathrm{ac}}=0$ and $C_{+}=33.8(0.2)$, showing both theoretical and experimental
isolation over $20\,\mathrm{dB}$ with a bandwidth of 20 MHz. Around $\Delta\sim0$, 
an isolation over $36.7\,\mathrm{dB}$ is expected from the theory, as the purple curve shows.
However, the background-corrected experimental data points 
around $\Delta=0$ are absent because the residual transmission 
of the blocked field is much weaker than the background, which makes the background-corrected value
inaccessible from the fluctuation of the background with the current data integration time of 100 $\mu$s. We believe the actual isolation around $\Delta=0$ should be comparable with the theoretical prediction.
We can also see that an isolation of approximately $-10\,\mathrm{dB}$ can also be obtained when the
probe is resonant to the polariton states. 

The large $\mathcal{I}$ around
zero detuning confirms the high performance of the optical isolator
based on the non-reciprocal polariton in our scheme. 
From Eq.$\,$(\ref{eq1}), when $\Delta_{\mathrm{ac}}=\Delta=0$,
the ideal isolation 
\begin{equation}
\mathcal{I}_{\mathrm{max}}=(1+2C_{+})^{2}\label{eq2}
\end{equation}
would be achieved. $\mathcal{I}_{\mathrm{max}}$ increases quadratically 
with the cooperativity $C_{+}$, and over 30-dB isolation can be achieved
as $C_{+}\geq15.3$.

We then comprehensively investigate the performance
of the isolator by varying the intracavity atom number 
($N_\mathrm{eff}$) and probe intensity under the condition $\Delta_{\mathrm{ac}}=\Delta=0$, 
and the results are summarized in Figure~\ref{fig3}.
The $\left(-z,\sigma_{-}\right)$ light couples to an empty cavity mode (Figure~\ref{fig1}d), and the 
transmission depends linearly on the input light power (Figure~\ref{fig3}a). 
The slight decrease in the transmission at larger
$C_{+}$ is because of the scattering of the atom with imperfections in state preparations.
An overall transmission of 18\% is obtained and is limited by the impedance 
mismatch of the FPC~\cite{Yang2019}. A much higher transmission
would be achieved by optimizing impedance matching of the cavity. 
For a given $C_{+}$, in Figure~\ref{fig3}b, the transmission of the $\left(+z,\sigma_{+}\right)$ 
light is blocked at weak input power with a corresponding mean photon number less than 0.1, as expected. 
However, the input field can also excite the polariton modes off-resonantly, which would degrade 
the isolation ratio as the input power increases with a fixed atom number. 
The polariton modes will be fully saturated at a certain input power which depends on the atom number.
The larger $C_{+}$ (also the larger atom number) is, the more difficult it is for the polariton modes to be excited and the better the isolation obtained.  
The dependence of the measured isolation $\mathcal{I}$ on $C_{+}$ compared with the theoretical curve (Eq.~\ref{eq2}) is shown in Figure~\ref{fig3}c. 
Although the measured $\mathcal{I}$ is slightly lower than the theoretical value
$\mathcal{I}_{\mathrm{max}}$ (red solid curve), which is mainly limited by the experimental imperfections, such as background noise and the finite fidelity of the target spin-polarized atomic state preparation as discussed 
above, the measured isolation can still reach 30 dB with $C_{+}$ of approximately 35. 

\subsection{Performance of the optical isolator with two opposite laser beams coexisting}

Different from many other magnet-free optical nonlinear non-reciprocal devices, the 
optical isolator demonstrated in this article can work in the condition with coexisting forward and backward light due to the decoupling of the two circularly polarized cavity modes. This property would dramatically expand the application. The isolation is experimentally verified with the input power varying in this bidirectional-probing scenario, and the data for $C_{+}=15.3(0.5)$ and $10.8(0.3)$ are shown by red circles and blue squares in Figure~\ref{fig3}d, respectively.
The behavior is quite similar to that of single-direction probing, where a higher $C_{+}$ gives higher isolation with weak input power.
However, because both the forward and backward lights 
are shined, it is very difficult to separate the transmission of the backward light from the residual reflection of the forward light, and the measured isolation is slightly lower. The actual isolation should be the same as the case with an individual probing light field.

\subsection{Reconfigurability of the optical isolator}

Moreover, our device is reconfigurable by controlling the internal state of spin-polarized atoms. This is done experimentally by manipulating the population of the atomic Zeeman states. To verify this, the circularity of the 459-nm laser is switched between $\sigma_{-}$ and $\sigma_{+}$ polarization with a time period of 15 ms to switch the atomic population between states $S_{z}=-4$ ($|6S_{1/2},F=4,m_{F}=-4\rangle$) and $S_{z}=4$ ($|6S_{1/2},F=4,m_{F}=+4\rangle$) back and forth. The blocking and transmitting directions are then
reconfigured accordingly (Figure~\ref{fig4}). Here, a constant 1.5-Gauss magnetic field is applied to maintain the spin polarization of atoms. We find that the shut-off time $43\pm2$ $\mu$s is much faster than the turn-on time $705\pm19$ $\mu$s. The difference between these two switching times is due to the different joint effects of optical pumping and atom-cavity coupling in the two processes. The detailed analysis can be found in the Supporting Information (Section S11). 
The theoretical curves of the switching calculated by taking both the optical pumping process and the 
atom-cavity couplings in all Zeeman states into account are also shown in Figure~\ref{fig4} (solid curves). The experimental results (open circles) are in good agreement with the theoretical prediction.

\subsection{Validity of the isolation without a magnetic field} 

\begin{figure}
\centering{}\includegraphics[width=1\columnwidth]{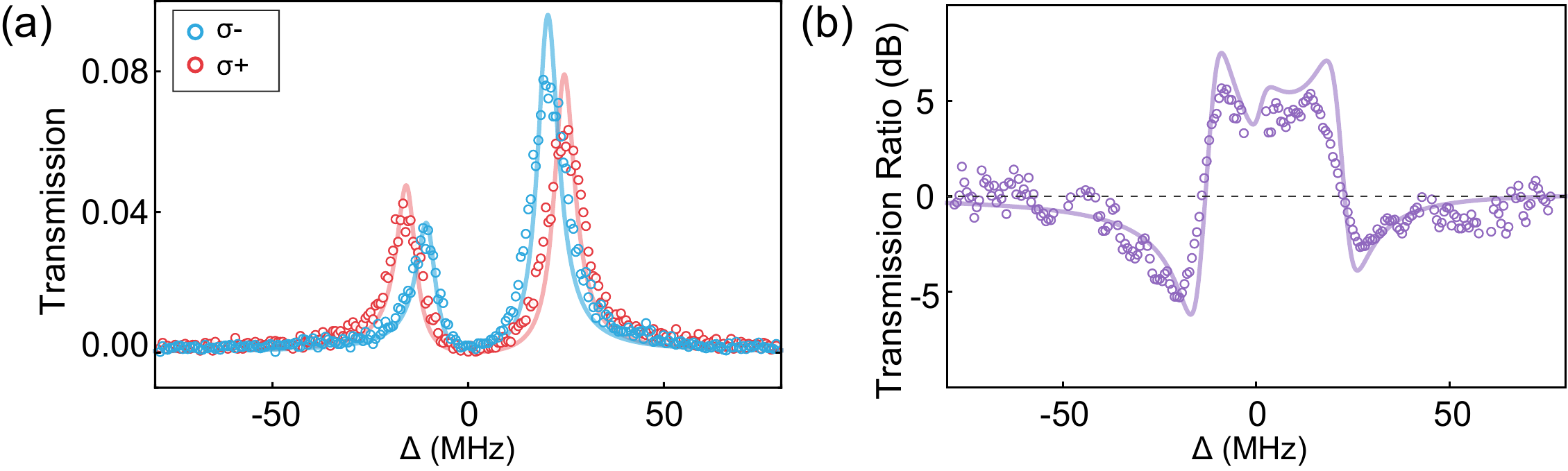} \caption{\label{fig5} Validity of the isolation without a magnetic field. a) The non-reciprocal spectra for the $\sigma_{+}$ (open red circles) and $\sigma_{-}$ (open blue circles) probes. The solid curves are the theoretical fittings by Eq.~\ref{eq1}. b) The transmission ratio with zero magnetic field, where a 4-dB isolation with bandwidth of 30 MHz is maintained.
The purple solid curve is the theoretical expectation from the two fitting curves in a).}
\end{figure}

The non-reciprocal cavity polariton and the demonstration of the aforementioned optical isolator are performed with a weak bias magnetic field.
Notably, even such a weak magnetic field is not necessary for our current device.
The performance of the isolation under a zero magnetic field is demonstrated and measured 
when the atoms are continuously pumped
with a $\sigma_{-}$-polarized 459 nm optical pumping laser, by which the spin direction of atoms can be preserved. The data are shown in Figure~\ref{fig5}. Here, the
atomic polarization is degraded due to the absence of the quantization axis previously defined by the magnetic field. However, the presence of the $\sigma_{-}$-polarized 459 nm pump laser can weakly maintain the polarization, and thus, the system still gives non-reciprocal spectra, as shown in Figure~\ref{fig5}a, where blue (red) open circles are for the $\sigma_{-}$ ($\sigma_{+}$) polarized probe. 
The corresponding isolation of the probe light fields is shown in Figure~\ref{fig5}b, which gives an isolation around 
$4\,\mathrm{dB}$ with a bandwidth over 30 MHz. It should be noted that by increasing the power of the polarized 459 nm optical pumping laser, a better isolation performance can be achieved.

\subsection{Non-reciprocal quantum statistics} 

\begin{figure}
\centering{}\includegraphics[width=0.8\columnwidth]{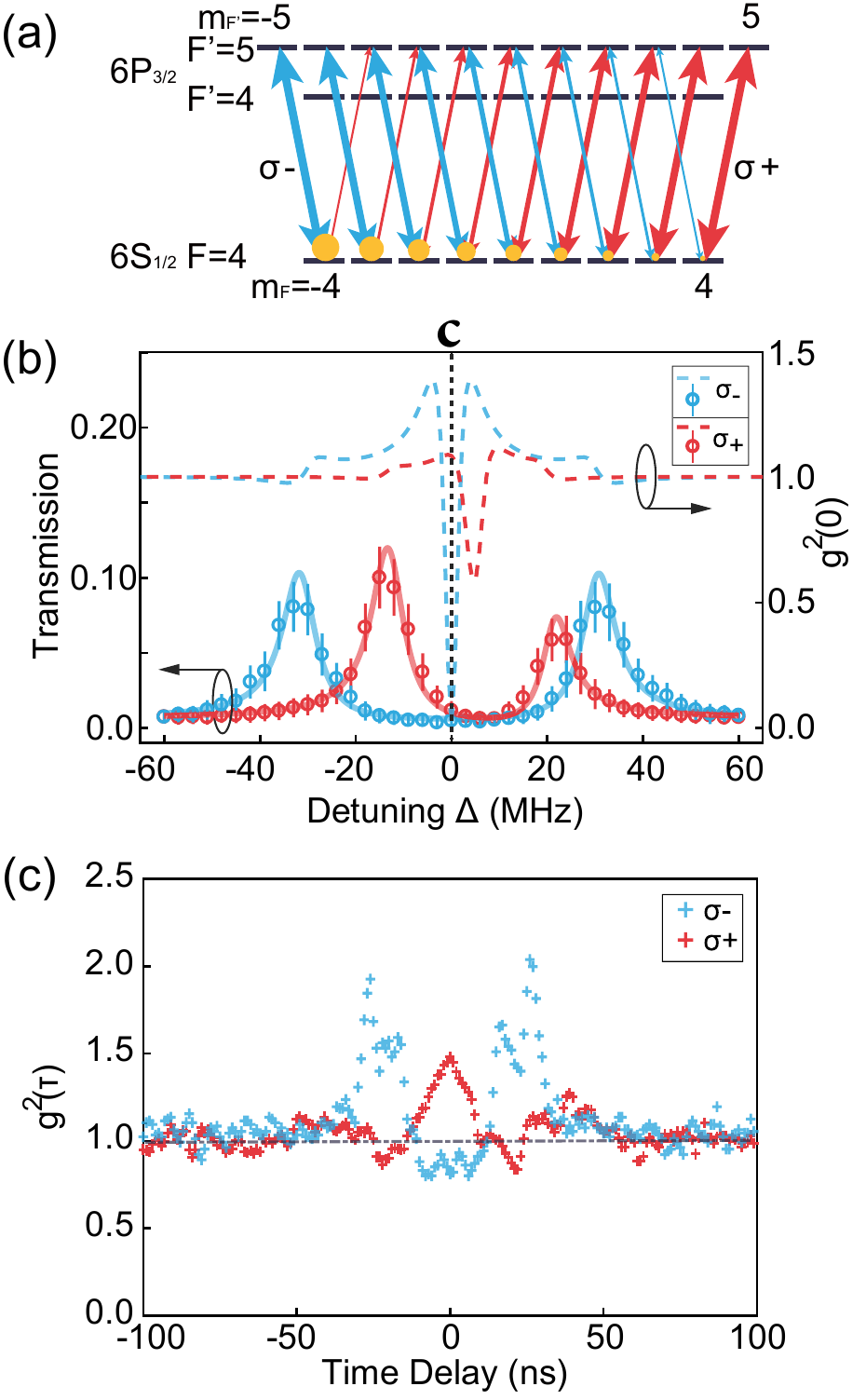} \caption{\label{fig6} Quantum interference of polaritons and the nonreciprocity of quantum statistics.
a) The energy diagram for realizing non-reciprocal polaritons
under more general conditions. The atom ground spin state has a slanted population distribution on $m_{F}$. The size of yellow solid circles indicates the population in every Zeeman sublevel. The thickness of the blue ($\sigma_{-}$) and red ($\sigma_{+}$) arrows represents the corresponding $\sigma$-transition strength. 
b) The measured spectra by probing the system with $\sigma_{+}$ (red open circles) and
$\sigma_{-}$ (blue open circles) polarization. The fitting of the data gives $C_{+}=15.1(0.3)$, 
$C_{-}=50.8(0.6)$, and $\Delta_{\text{ac}}=0$ for
the $\sigma_{-}$ mode and $\Delta_{\text{ac}}=6\,\mathrm{MHz}$ for
the $\sigma_{+}$ mode. The theoretical expectations of the second-order
correlation function $g^{(2)}(0)$ are displayed as dashed curves,
and the nonreciprocity of quantum statistics is clearly presented.
c) The measured temporal second-order correlation function
$g^{(2)}(\tau)$ by detecting the emitted photons from the cavity with the probe-cavity detuning fixed at $0\,\mathrm{MHz}$. The error bars
correspond to one standard deviation.}
\end{figure}

The non-reciprocal polariton is then studied in more general cases since the maximally polarized spin state might not always be available in experiments.
For example, as schematically illustrated in Figure~\ref{fig6}a, both the $\sigma_{+}$- and $\sigma_{-}$-modes of the cavity could strongly
couple to the atomic transition with unequal strengths ($C_{\pm}\gg0$
and $C_{-}\ne C_{+}$). 
The situation is realized by preparing the atom to state $|6S_{1/2},F=4\rangle$ with a slanted population on the Zeeman sublevels from $m_{F}=-4$
to $m_{F}=4$ (yellow solid circles). 
The cavity is then tuned to be resonant to $|6S_{1/2},F=4\rangle\leftrightarrow|6P_{3/2},F'=5\rangle$.
Owing to the unequal transition strength (arrows with different thicknesses in Figure~\ref{fig6}a), not only is the frequency of the polaritons asymmetric, but the strength of the nonlinearity experienced by the polariton is also different. The spectra of the polariton probed
by weak coherent light, shown in Figure~\ref{fig6}b, verify the nonreciprocity of the polariton. Here, the system is probed with $\sigma_{+}$- and $\sigma_{-}$-polarized light both in the $(+z)$ direction, which is equivalent to the case in which the system is probed by the $(+z, \sigma_{+})$- and $(-z, \sigma_{-})$-probing configurations. 
The spectra of two $\sigma_{+}$-polaritons
are asymmetric due to the off-resonant coupling
between the bare cavity modes and other hyperfine energy levels ($|6P_{3/2},F'=4\rangle$)
of the atoms, which induces a 6-$\mathrm{MHz}$ frequency shift in the effective
cavity mode. In particular, such a dispersive effect
is also non-reciprocal and is similar to the avoid-crossing results
for large atom-cavity detuning in Figure~\ref{fig2}d.

In such a few-atom cavity QED system, the anharmonicity of the polariton is inversely proportional to $\sqrt{N_{\mathrm{eff}}}$ (Section S5 in the Supporting Information). The quantum effects of the polariton are greatly suppressed because the Kerr coefficient is much smaller than the system dissipation rate ($g_{\mathrm{eff}}/2N_{\mathrm{eff}}\ll\kappa,\gamma$). Surprisingly, we theoretically predict that the transmitted light exhibits remarkable quantum statistics for a classical input, as shown by the second-order correlation $g^{(2)}(0)$ in Figure~\ref{fig6}b (blue and red dashed lines). When the probe is near resonant to the polariton states, the emissions of both $\sigma_{\pm}$-polaritons show a slight deviation from the Poissonian statistic ($g^{(2)}(0)=1$) due to the weak anharmonicity of the polaritons. In contrast, much more pronounced nonreciprocity of the quantum statistics occurs around $\Delta=0$, where the super-Poissonian distribution ($g_{\sigma_{+}}^{(2)}(0)>1$) versus the sub-Poissonian distribution ($g_{\sigma_{-}}^{(2)}(0)<1$) is expected (indicated by the vertical black-dotted line with mark ``$\mathbf{c}$'' in Figure~\ref{fig6}b). The physical mechanism behind the extraordinary behavior with $\Delta\sim0$ is the quantum interference between the two polariton states $p_{\pm}$, which possesses a Kerr coefficient with opposite signs (Section S5 in the Supporting Information).

The experimental results of the temporal second-order correlation
functions $g^{(2)}(\tau)$ for the $\sigma_{\pm}$-modes at $\Delta=0$ are shown in Figure~\ref{fig6}c, where the red (blue) cross markers are for the $\sigma_{+}$- ($\sigma_{-}$-) mode. Here, we investigate the nonclassical
statistics of the emission from cavity polaritons excited by a forwarding probe beam with fixed power. For the $\sigma_{+}$-polariton, the probe frequency is actually
closer to one polariton state and shows a super-Poissonian distribution
($g^{(2)}(0)\sim1.5$) and bunching effect ($g^{(2)}(0)>g^{(2)}(\tau)$).
In contrast, the probe is equally off-resonant to both $\sigma_{-}$-polaritons, where significant destructive interference of cavity polaritons takes place. The sub-Poissonian distribution ($g^{(2)}(0)\sim0.8$) and
anti-bunching effect ($g^{(2)}(0)<g^{(2)}(\tau)$) are displayed (Section S6 in the Supporting Information). The deviation between the experimental and theoretical results may be attributed to the noise background and system parameter uncertainties. Additionally, there are substantial oscillations in the temporal correlation function for both $\sigma_{\pm}$-polaritons, which indicate quantum interference between different polariton states. These results suggest that the anharmonicity of polaritons would bring non-reciprocal quantum features, which would extend the concept of non-reciprocal devices to the quantum regime~\cite{Huang2018} and provide experimental insights into the unconventional bosonic blockade effect~\cite{Snijders2018}.

\section{Conclusion}
Our demonstration of non-reciprocal polaritons from both linear and
nonlinear aspects opens up a new perspective for research on
cavity QED as well as novel non-reciprocal devices for photonics.
The collective effect of a small atom ensemble
produces polariton states as a hybridization of light and matter.
By manipulating the atom at a maximally polarized state, the polariton can only be accessed optically from a certain direction, allowing isolation of photons with an isolation ratio exceeding $30\,\mathrm{dB}$ on the single-photon level. The reconfigurability of the isolation directions is experimentally
demonstrated by switching the polarization of the atoms. The validity of the isolator with zero magnetic
field is verified with relatively low isolation. 
By manipulating the quantum interference between the polaritons, the weak anharmonicity could still generate non-reciprocal nonclassical outputs from classical input. With the direction-dependent bunching and anti-bunching properties of polaritons being observed for the first time, our experiments show the potential for realizing both linear
and nonlinear non-reciprocal optics effects on the single-quanta level.
Such quantum non-reciprocal polariton states can be extended
to phonons~\cite{Golter2016} and microwave photons~\cite{Putz2014,Bienfait2016}
by harnessing their coupling with electron spin ensembles, and could
find applications as quantum routers and isolators for quantum networks~\cite{Lodahl2017}.
Our work also initiates the exploration of the concept of quantum
nonreciprocity with conventional quantum optics systems from two
perspectives: on the one hand, polaritons are inherently a hybrid quantum state of photons and atoms, so superpositions of reciprocal and non-reciprocal polariton states are worth further investigation. On
the other hand, we can study the quantum behaviors of non-reciprocal
polaritons utilizing their intrinsic anharmonicity.




\medskip
\noindent \textbf{Supporting Information} \\ 
Supporting Information is available from the author.

\medskip
\noindent \textbf{Acknowledgements}\\ We thank H J. Kimble for helpful discussions. We also thank Ya-Nan Lv for her help in the numerical simulations.This work was supported by the National Key Research and Development Program of China (Grant Nos. 2021YFA1402002 and 2017YFA0304502), the National Natural Science Foundation of China (Grant No. U21A6006, U21A20433, 11974223, 11974225, 12104277, and 12104278), and the Fund for Shanxi 1331 Project Key Subjects Construction. M.L. and C.-L.Z. were supported by National Natural Science Foundation of China (Grant No.11674342 and 11922411), and  the Program of State Key Laboratory of Quantum Optics and Quantum Optics Devices.

\medskip
\noindent \textbf{Conflict of interest}\\
The authors declare no competing interests.

\medskip
\noindent \textbf{Data Availability Statement}\\
The data that support the findings of this study are available from the corresponding author upon reasonable request.
\medskip

%




\bibliographystyle{Zou}
\bibliography{reference}

\pagebreak

\clearpage

\widetext

\renewcommand{\thefigure}{S\arabic{figure}}

\renewcommand{\theequation}{S\arabic{equation}}
\setcounter{equation}{0} 
\setcounter{figure}{0}

\section{Supplementary information for: Non-reciprocal Cavity Polariton with Atoms Strongly Coupled to Optical Cavity}
\begin{noindent}
\noindent{{$\text{Pengfei~Yang}^{1,3,4,\dag}, \text{Ming Li}^{2,\dag}, \text{Xing Han}^{1,3}, \text{Hai He}^{1,3}, \text{Gang Li}^{1,3,*}$,$ \text{Chang-Ling~Zou}^{2,1,*}$, $\text{Pengfei Zhang}^{1,3}$, $\text{Yuhua Qian}^{4,5}$ and  
 $\text{Tiancai Zhang}^{1,3,*}$}
}

\noindent{$^{1}$State Key Laboratory of Quantum Optics and Quantum Optics Devices, and Institute of Opto-Electronics, Shanxi University, Taiyuan 030006, China.}\\
\noindent{$^{2}$CAS Key Laboratory of Quantum Information, University of Science and Technology of China, Hefei 230026, China.}\\
\noindent{$^{3}$Collaborative Innovation Center of Extreme Optics, Shanxi University, Taiyuan 030006, China.}\\
\noindent{$^{4}$Institute of Big Data Science and Industry, Shanxi University, Taiyuan 030006, China.}\\
\noindent{$^{5}$Key Laboratory of Computational Intelligence and Chinese Information Processing of Ministry of Education, Shanxi University, Taiyuan 030006, China.}\\

\noindent{\textbf{Corresponding author E-mail:}\\Gang Li, gangli@sxu.edu.cn\\Chang-Ling~Zou, clzou321@ustc.edu.cn\\Tiancai Zhang, tczhang@sxu.edu.cn}

\end{noindent}
\clearpage

\noindent The supplementary material presents a theoretical analysis of non-reciprocal cavity polaritons and single-photon-level optical isolation, simulation of the reconfiguration of the isolator, experimental setups and time sequences.

\vbox{}

\begin{figure}[htbp]
\centering
\includegraphics[width=1 \linewidth]{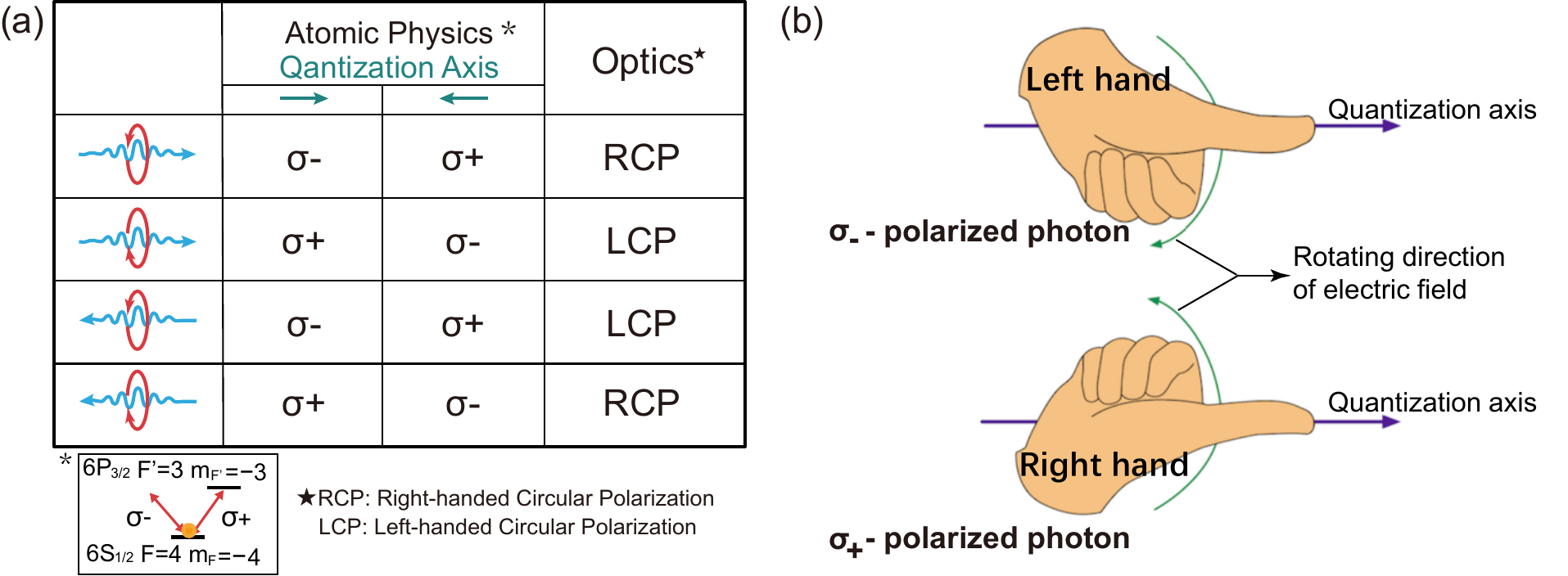}
\caption{Circularly polarized light. (a) Descriptions of circularly polarized light in atomic physics and optics. The first row displays the rotation of the electric field and the propagation directions for a light field. The second row is the classification of the polarization in atomic physics, where a quantization axis is referenced. The third row gives the classification of the polarization in optics. The small inset shows the energy levels and corresponding $\sigma$ transitions we used to demonstrate the non-reciprocal cavity polaritons and the single-photon-level isolator in the maintext. (b) Depictions of the chiral photons where $\sigma_{-}$-polarized and $\sigma_{+}$-polarized photons are described by the left and right hands, respectively.}
\label{fig:S1}
\end{figure}

\begin{center}
\textbf{\large S1. Circularly polarized light in atomic physics and optics}\\
\end{center}
\noindent For circularly polarized light, the electric-field vector rotates around the wave propagation direction. 
The definitions of circular polarization are different in atomic physics and optics. The relations of the circularly polarized light fields with definitions in atomic physics and optics are summarized in Figure \ref{fig:S1}(a). 

In optics, the polarization is defined by the rotation direction of the electric
field by looking against the propagation direction of the light beam.
The polarization is defined as left-handed circular polarization (LCP) for a counterclockwise rotation of the electric field. The right-handed circular polarization (RCP) is for a clockwise rotation of the electric field. 

In atomic physics, however, the polarization of light is defined by the rotation direction of the electric field regardless of the propagating direction but with respect to a quantization direction. The direction is often provided by a magnetic field. In this scenario, $\sigma_{-}$- ($\sigma_{+}$-) polarized light is light with its electric field rotating clockwise (counterclockwise) if viewed against the quantized direction. When interacting with an atomic dipole, the $\sigma_{-}$- ($\sigma_{+}$-) polarized light drives atomic transition with a change in the Zeeman quantum number $\Delta m_F = -1 \ (+1)$. We also find that the two circularly polarized photons defined in atomic physics are chiral photons. The depictions of the chiral photons are displayed in Figure \ref{fig:S1}(b), in which the $\sigma_{-}$-polarized and $\sigma_{+}$-polarized photons are well described by the left and right hands, respectively.

In our experiment demonstrating the chiral interaction and the optical isolator, $\sigma_{+}$-polarized light drives the transition from $|6S_{1/2},F=4,m_{F}=-4\rangle$ to $|6P_{3/2},F'=3,m_{F}=-3\rangle$, and $\sigma_{-}$-polarized light does not interact with the atom due to the absence of the corresponding Zeeman sublevel $m_{F}=-5$ in the excited state $|6P_{3/2},F'=3\rangle$. The energy level diagram and the corresponding interactions are shown in the inset of Figure \ref{fig:S1}(a).  

In the experiment demonstrating the non-reciprocal quantum statistics, state $|6P_{3/2},F'=5\rangle$ is adopted as the excited state. In this case, the $\sigma_{-}$- ($\sigma_{+}$-) polarized light drives atomic transition with the change in Zeeman quantum number $\Delta m_F = -1 \ (+1)$ between states $|6S_{1/2},F=4\rangle$ and $|6P_{3/2},F'=5\rangle$. The transition strengths are different for the Zeeman state with $m_F=-4$ to $+4$ in state $|6S_{1/2},F=4\rangle$, which can be found in the Cesium D Line Data \cite{Daniel:2019}.

\begin{figure}[htbp]
\centering
\includegraphics[width=.6\linewidth]{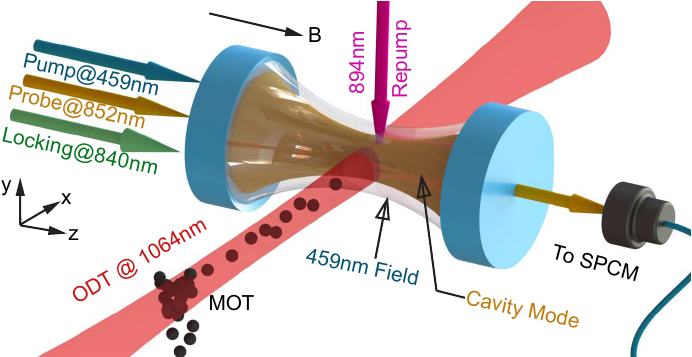}
\caption{Experimental
apparatus. A high-finesse Fabry-P{\'e}rot cavity is horizontally oriented,
and an MOT is located $\sim8\,\mathrm{mm}$ beside the cavity mode
(horizontal). The laser beam of the ODT at $1064\,\mathrm{nm}$ is utilized
to transport atoms from the MOT into the cavity mode. A weak probe beam along the
axis of the cavity and SPCMs are implemented to record the transmission
of the cavity.}
\label{fig:S2}
\end{figure}

\begin{center}
\textbf{\large S2. Experimental setups and time sequences}\\
\end{center}
\noindent As shown in Figure \ref{fig:S2}, the experimental
setup consists of a horizontally oriented high-finesse Fabry-P{\'e}rot
cavity with a length of $335\,\mathrm{\mu m}$, and the $\mathrm{TEM_{00}}$
mode has a waist of $33.3\,\mathrm{\mu m}$. A $1064\,\mathrm{nm}$
optical dipole trap (ODT) laser beam (horizontal) with a waist of
$36\:\mathrm{\mu m}$ is used to load cold atoms from the MOT and transfer the 
atoms to the cavity. The beam direction is perpendicular to the cavity axis. 
The output of the cavity is the emission of the cavity polaritons or bare cavity modes, is recorded by single photon counting modules (SPCMs). The parameters of the cavity quantum electrodynamics (QED) system are $(g_{0},\kappa,\gamma)=2\pi\times(1.7,3.7,2.6)\,\mathrm{MHz}$,
where $g_{0}$ is the maximal single atom-cavity coupling strength for the
$\sigma_{+}$ transition of $^{133}\text{Cs}$ $|6S_{1/2},F=4,m_{F}=-4\rangle\leftrightarrow|6P_{3/2},F'=3,m_{F'}=-3\rangle$,
$\kappa$ is the cavity field decay rate, and $\gamma$ is the atomic
excitation decay rate. 
An auxiliary $840\,\mathrm{nm}$ laser beam
along the cavity axis is used to stabilize and control the length of the cavity.
A $459\,\mathrm{nm}$ laser with a waist of $550\,\mathrm{\mu m}$ oriented along the cavity axis and a $894\,\mathrm{nm}$ repump laser
beam perpendicular to the cavity axis are used to prepare the atom
in the Zeeman state $|6S_{1/2},F=4,m_{F}=-4\rangle$.

\begin{figure}[htbp]
\centering
\includegraphics[width=.8\linewidth]{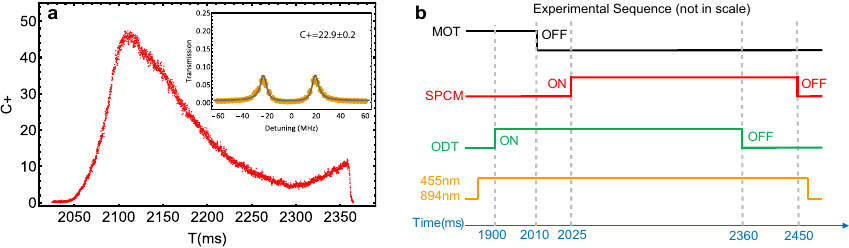}
\caption{ Determination of $C_{\pm}$ and experimental sequences. (\textbf{a}) The collective atom-cavity coupling cooperativity
$C_{+}$ during measurements. The red
dots indicate the $C_{+}$ for each moment during the experiment, and they are 
extracted from atom transportation 600 times. $C_{+}$ is
extracted from the transmission spectra by fitting the vacuum Rabi splitting 
spectrum, as shown in the inset figure, where $C_{+}=22.9\pm0.2$. 
$C_{-}$ is determined by the samiliar method. In our 
experiments, the cold atoms move very slowly ($\sim\mathrm{cm/s}$). As a result, 
although the atom number inside the cavity varies with time, the measurement of 
the system with duration on the ms-level reflects the instantaneous state of the 
system. (\textbf{b}) shows the detailed time sequence of the experiment. The MOT
is open from 0 to 2010 ms for atom loading. The ODT is applied at 1900 ms and 
is kept for 460 ms until the atomic assemble transit through
the cavity mode. The 455 nm and 894 nm laser beams are kept on throughout the measurement to ensure that the atoms are always in state $|6S_{1/2},F=4,m_{F}=-4\rangle$. 
The SPCM records the transmission of the cavity from 2025 ms to 2450 ms.}
\label{fig:S3}
\end{figure}

Figure \ref{fig:S3}a shows the variance of the parameter of cooperativity $C_+$
versus time during the experiment when the atoms flow from the MOT to the
cavity mode. $C_+$ is deduced from the vacuum Rabi splitting, as shown
in the inset of the figure. 
The time sequences for the experiment are shown in Figure \ref{fig:S3}b. 
Cold atoms are initially accumulated in a mirror-reflected magneto-optical trap
(MOT) from the background atomic vapor, and the atom cloud is approximately $8\,\mathrm{mm}$ away from the cavity center. An additional 5$\,\mathrm{ms}$ polarization gradient cooling procedure is performed to precool the atoms, and the
repumping laser of the MOT is kept at $2\,\mathrm{ms}$ to pump atoms
into state $|6S_{1/2},F=4\rangle$. The ODT and optical pumping lasers
are switched on before shuting off the MOT lasers. Since the ODT is aligned
to overlap with the MOT and the waist is set at the right middle of the
MOT and the cavity, cold atoms can be guided and successively transported
into the cavity by the ODT after shuting off the MOT. The MOT
is kept from 0-$2010\,\mathrm{ms}$ for atom loading, during which an ODT
is applied at $1900\,\mathrm{ms}$ and is kept for $460\,\mathrm{ms}$
until the atomic assemble flows out of the cavity mode. The $459\,\mathrm{nm}$
and $894\,\mathrm{nm}$ laser beams are kept on throughout the measurement
to prepare atoms in state $|6S_{1/2},F=4,mF=-4\rangle$ continuously.
The single photon counting module (SPCM) records the transmitted photon of the cavity from $2025\,\mathrm{ms}$
to $2450\,\mathrm{ms}$ to characterize the non-reciprocal polaritons
in the few-atom cavity QED system.

\vbox{}

\begin{center}
\textbf{\large S3. Model}\\
\end{center}
\noindent For our experimental setup, the atoms pass through the cavity field
with a duration of $\sim1\,\mathrm{ms}$, which is orders longer
than the relaxation times of the atom and cavity; thus, we treat the system
as steady state at any instance. The Hamiltonian that describes the
coupling between a single-mode field and $N$ two-level atoms is 
\begin{eqnarray}
H & = & \Delta a^{\dagger}a+\frac{\Delta'}{2}\sum_{i=1}^{N}s_{z}^{i}+\sum_{i=1}^{N}g_{i}\left(a^{\dagger}s_{-}^{i}+as_{+}^{i}\right)\label{eq:cqed}
\end{eqnarray}
where $a$ ($a^{\dagger}$) is the annihilation (creation) operator
of the cavity field, $\Delta'=\Delta+\Delta_{ac}$, $\Delta$ is the
probe-cavity detuning, $\Delta_{ac}$ is the atom-cavity detuning,
$s_{-}^{i}$and $s_{+}^{i}$ represent the lowering and raising operators
of the transition of the $i$-th atom and fulfill $\left[s_{+}^{i},s_{-}^{i}\right]=s_{z}$,
and $g_{i}$ is the atom-cavity strength for the $i$-th atom. Here, the
full model is very difficult to solve in practice when $N>15$, due
to the high computational complexity. By approximately treating the
coupling between a few atom ensemble and the cavity as an $N$-atom
ensemble uniformly coupled with the cavity, i.e., $g_{i}=g$, the collective
behavior of the $N$ atoms can be treated as a $N/2$ spin with operators
\begin{eqnarray}
L_{+} & = & \sum_{i=1}^{N}s_{+}^{i}\\
L_{-} & = & \sum_{i=1}^{N}s_{-}^{i}\\
L_{z} & = & \sum_{i=1}^{N}s_{z}^{i}
\end{eqnarray}
It is easy to verify that $\left[L_{+},L_{-}\right]=\sum_{i}^{N}\left[s_{+}^{i},s_{-}^{i}\right]=L_{z}$.
The spin operator can be transformed to a bosonic operator by the Holstein-Primakoff
transformation,
\begin{eqnarray}
L_{+} & = & \sqrt{N}b^{\dagger}\sqrt{1-\frac{b^{\dagger}b}{N}}\\
L_{-} & = & \sqrt{N}\sqrt{1-\frac{b^{\dagger}b}{N}}b\\
L_{z} & = & \left[L_{+},L_{-}\right]=-N+2b^{\dagger}b
\end{eqnarray}
Neglecting the constant term $-N$, the Hamiltonian is transformed
to
\begin{eqnarray}
H & = & \Delta a^{\dagger}a+\Delta'b^{\dagger}b+g\left(a^{\dagger}L_{-}+aL_{+}\right)\nonumber \\
 & = & \Delta a^{\dagger}a+\Delta'b^{\dagger}b+g\sqrt{N}\left[\sqrt{1-\frac{b^{\dagger}b}{N}}a^{\dagger}b+ab^{\dagger}\sqrt{1-\frac{b^{\dagger}b}{N}}\right],
\end{eqnarray}
which describes the interaction between two bosonic modes. The square
root of the bosonic operator can be expanded in Taylor series as
\begin{equation}
\sqrt{1-\frac{b^{\dagger}b}{N}}=1-\frac{b^{\dagger}b}{2N}-\frac{\left(b^{\dagger}b\right)^{2}}{8N^{2}}-...\label{eq:Taylor}
\end{equation}
In the following analysis, we assume that the system is probed with a low
excitation level, i.e., the excitation $\langle b^{\dagger}b\rangle\ll N$,
only the first few terms should be taken into consideration. For zero-th-
and first-order approximations, we will separately discuss the linear and
nonlinear regimes in the following sections.

\vbox{}

\begin{center}
\noindent \textbf{\large S4. Scattering matrix in the linear regime}\\
\end{center}

\noindent For a low excitation $\frac{\langle b^{\dagger}b\rangle}{N}\ll1$,
and applying the linear approximation $\sqrt{1-\frac{b^{\dagger}b}{N}}\approx1$
{[}Eq.~(\ref{eq:Taylor}){]}, the Hamiltonian can be simplified to
the form of two linearly coupled bosonic modes
\begin{eqnarray}
H_{l} & = & \Delta a^{\dagger}a+\Delta'b^{\dagger}b+g_{\mathrm{eff}}\left(a^{\dagger}b+ab^{\dagger}\right).
\end{eqnarray}
Due to the collectively enhanced effective coupling strength $g_{\mathrm{eff}}=g\sqrt{N}$,
the cavity mode and the collective atom spin are hybridized and constitute
the atom-cavity polariton states $p_{u}$ and $p_{l}$. They are linear
combinations of mode $a,b$ and can be derived by solving the eigenvector
of the matrix
\begin{eqnarray}
\mathrm{H} & = & \left(\begin{array}{cc}
\Delta & g_{\mathrm{eff}}\\
g_{\mathrm{eff}} & \Delta'
\end{array}\right).
\end{eqnarray}
The Hamiltonian can be diagonalized to $H=\Delta_{u}p_{u}^{\dagger}p_{u}+\Delta_{l}p_{l}^{\dagger}p_{l}$,
with the eigenfrequencies of the cavity polaritons being
\begin{eqnarray}
\Delta_{u,l} & = & \Delta+\frac{\Delta_{ac}\pm\sqrt{\Delta_{ac}^{2}+4g_{\mathrm{eff}}^{2}}}{2}.\label{eq:d+}
\end{eqnarray}
For $g_{\mathrm{eff}}>\kappa,\gamma$, typical anti-crossing spectra
should be observed.

\begin{figure}[htbp]
\centering
\includegraphics[width=.6\linewidth]{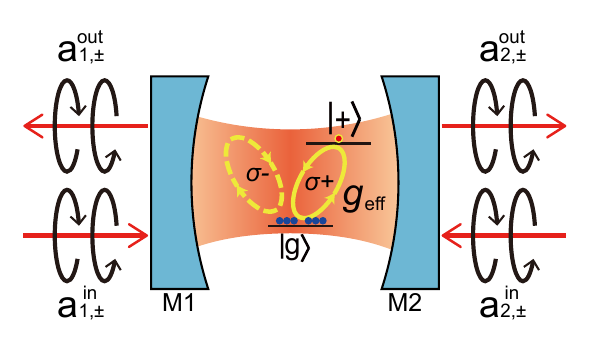}
\caption{Input and output modes at two different F-P cavity mirrors. The subscript $1$ ($2$) denotes the mode at the left (right)
mirror, and $+$ ($-$) denotes the right-handed (left-handed) polarization.}
\label{fig:S4}
\end{figure}

\noindent In our experiment, there are $\sigma_{+}$- and $\sigma_{-}$-polarized
modes ($b_{\pm}$) coupling with atoms. Due to the population of atoms
on hyperfine ground states, we have the effective coupling strength
between atoms to the two modes are orthogonal and different. Therefore,
we introduce effective coupling strengths $g_{\mathrm{eff},\pm}$ for the coupling strength weighted by the Clebsch-Gordan coefficients
and population distributions. For a probe field with strength of $\varepsilon_{p}$
, i.e., 
\begin{equation}
H_{p}=\sqrt{2\kappa_{1}}\varepsilon_{p}\left(a+a^{\dagger}\right),\label{eq:Probe}
\end{equation}
the dynamics of the system follows
\begin{eqnarray}
\frac{d}{dt}a & = & \left(-i\Delta-\kappa\right)a-ig_{\mathrm{eff,\pm}}b_{\pm}+\sqrt{2\kappa_{1}}\varepsilon_{p},\\
\frac{d}{dt}b_{\pm} & = & \left(-i\Delta'-\gamma\right)b_{\pm}-ig_{\mathrm{eff,\pm}}a,
\end{eqnarray}
where $\kappa$ is the cavity decay rate, $\kappa_{1,2}$ is the coupling
strength to the cavity modes through mirrors $1$ and $2$, and $\gamma$
is the atom decay rate. The steady-state cavity field is obtained at $\frac{d}{dt}a=\frac{d}{dt}b_{\pm}=0$.
By the input-output formalism, the transmittance is derived as
\begin{eqnarray}
T_{\pm} & = & \frac{4\kappa_{1}\kappa_{2}}{{\kappa}^{2}}\left|\frac{1}{i\Delta/\kappa+1+2C_{\pm}/\left(i\left(\Delta+\Delta_{ac}\right)/\gamma+1\right)}\right|^{2},
\end{eqnarray}
where $C_{\pm}=g_{\mathrm{eff},\pm}^{2}/{2\kappa\gamma}$
is the cooperativity. For all atoms prepared to the Zeeman sublevel, coupling with $\sigma_{-}$ light is forbidden, i.e., $C_{-}=0$,
the maximum isolation ratio 
\begin{eqnarray}
\mathcal{I}_{0} & = & \left(1+2C_{+}\right)^{2}
\end{eqnarray}
could be achieved for $\Delta=\Delta_{ac}=0$.

To fully characterize the nonreciprocal property of the system, we
can investigate the scattering matrix that links the input and output
ports. Considering the two input/output modes at two different cavity
mirrors and two polarization states of each mode (as shown in Figure \ref{fig:S4}), the input-output
relationship is described by
\begin{eqnarray}
	\left(\begin{array}{c}
		a_{1,+}^{\mathrm{out}}\\
		a_{1,-}^{\mathrm{out}}\\
		a_{2,+}^{\mathrm{out}}\\
		a_{2,-}^{\mathrm{out}}
	\end{array}\right) & = & \mathbf{S}\cdot\left(\begin{array}{c}
		a_{1,+}^{\mathrm{in}}\\
		a_{1,-}^{\mathrm{in}}\\
		a_{2,+}^{\mathrm{in}}\\
		a_{2,-}^{\mathrm{in}}
	\end{array}\right),
\end{eqnarray}
where the subscript $1$ ($2$) denotes the mode at the left (right)
mirror and $+$ ($-$) denotes the right-handed (left-handed) polarization,
and the scattering matrix is
\begin{eqnarray}
	\mathbf{S} & = & \left(\begin{array}{cccc}
		0 & 1-\sqrt{\frac{\kappa_{1}}{\kappa_{2}}}t_{-} & t_{+} & 0\\
		1-\sqrt{\frac{\kappa_{1}}{\kappa_{2}}}t_{+} & 0 & 0 & t_{-}\\
		t_{+} & 0 & 0 & 1-\sqrt{\frac{\kappa_{2}}{\kappa_{1}}}t_{-}\\
		0 & t_{-} & 1-\sqrt{\frac{\kappa_{2}}{\kappa_{1}}}t_{+} & 0
	\end{array}\right),
\end{eqnarray}
where
\begin{eqnarray}
	t_{\pm} & = & \frac{2\sqrt{\kappa_{1}\kappa_{2}}}{\kappa}\frac{1}{i\Delta/\kappa+1+2C_{\pm}/(i(\Delta+\Delta_{ac})/\gamma+1)}.
\end{eqnarray}
The scattering matrix for intensity is 
\begin{eqnarray}
\mathbf{S_{I}} & = & \left(\begin{array}{cccc}
		0 & \left(1-\sqrt{\frac{\kappa_{1}}{\kappa_{2}}T_{-}}\right)^{2} & T_{+} & 0\\
		\left(1-\sqrt{\frac{\kappa_{1}}{\kappa_{2}}T_{+}}\right)^{2} & 0 & 0 & T_{-}\\
		T_{+} & 0 & 0 & \left(1-\sqrt{\frac{\kappa_{1}}{\kappa_{2}}T_{-}}\right)^{2}\\
		0 & T_{-} & \left(1-\sqrt{\frac{\kappa_{1}}{\kappa_{2}}T_{+}}\right)^{2} & 0
	\end{array}\right)
\end{eqnarray}
with $T_{\pm}$ the same as that in Eq. S16. For a reciprocal system, the scattering matrix fulfills $\mathbf{S}=\mathbf{S}^{T}(\mathbf{S_{I}}=\mathbf{S_{I}}^{T})$. For our system, since $C_{+}\neq C_{-}$, it can be directly inferred
that $\mathbf{S}\neq\mathbf{S}^{T}(\mathbf{S_{I}}\neq\mathbf{S_{I}}^{T})$ and the reciprocity is broken
by the different atom-cavity coupling strengths. At the experimental condition with $C_{+}\gg 1$, $C_{-}= 0$ and $\Delta=\Delta_{\text{ac}}=0$, $T_{+}\approx 0$ and $T_{-}= \frac{4\kappa_{1}\kappa_{2}}{\kappa^{2}}$, the matrix can be simplified as
\begin{eqnarray}
\mathbf{S_{I}} & = & \left(\begin{array}{cccc}
		0 & \left(1-\frac{\kappa_{1}}\kappa\right)^{2} & 0 & 0\\
		1 & 0 & 0 & T_{-}\\
		0 & 0 & 0 & \left(1-\frac{\kappa_{2}}\kappa\right)^{2}\\
		0 & T_{-} & 1 & 0
	\end{array}\right),
\end{eqnarray}
and thus breaks Lorentz reciprocity. 

If we only consider transmitting paths shown as a1 to b1 and a2 to b2 in Figure 2a in the main text and omit the reflections of the PBSs, the whole device can be seen as a black box. Both the input and output are horizontally polarized; thus, the scattering matrix is a two-port matrix. The scattering matrix can be expressed as $\mathbf{S_{I}}  = \left(\begin{matrix}
		0 & T_{+}\\
		T_{-} & 0  
	\end{matrix}\right)$ and we have $\left(\begin{matrix}
		b_{1}\\
		b_{2}    
	\end{matrix}\right)  = \mathbf{S_{I}}\left(\begin{matrix}
		a_{1}\\
		a_{2}  
	\end{matrix}\right)$. Under the experimental conditions mentioned above, the scattering matrix is then $\left(\begin{matrix}
		0 & 0\\
		\frac{4\kappa_{1}\kappa_{2}}{\kappa^{2}} & 0  
	\end{matrix}\right)$ and the Lorentz reciprocity is broken.
\vbox{}

\begin{center}
\noindent \textbf{\large S5. Individual cavity polaritons under coherent driving}\\
\end{center}

\noindent For the first-order approximation of the Taylor series {[}Eq.~(\ref{eq:Taylor}){]},
the nonlinear term 
\begin{eqnarray}
H_{nl} & = & -\frac{g_{\mathrm{eff}}}{2N}\left(a^{\dagger}b^{\dagger}bb+ab^{\dagger}b^{\dagger}b\right)
\end{eqnarray}
should be considered, which involves the four-excitation interaction
between modes $a$ and $b$. In the picture of polaritons, the Hamiltonian
is written as
\begin{eqnarray*}
H & = & \Delta_{u}p_{u}^{\dagger}p_{u}+\Delta_{l}p_{l}^{\dagger}p_{l}+\sum_{ijkm}\eta_{ijkm}p_{i}^{\dagger}p_{j}^{\dagger}p_{k}p_{m}+h.c.,
\end{eqnarray*}
where $i,j,k,m\in\{u,l\}$ and $\eta_{ijkm}$ is the nonlinear interaction
strength, $p_{u,l}=\frac{(a\pm b)}{\sqrt{2}}$. For simplicity, we consider the case in which the cavity and
atom are both resonant with the driving field $\Delta=\Delta'=0$,
and the Hamiltonian reduces to
\begin{eqnarray}
H_{\mathrm{eff}} & = & g_{\mathrm{eff}}p_{u}^{\dagger}p_{u}-g_{\mathrm{eff}}p_{l}^{\dagger}p_{l}\nonumber \\
 &  & -\frac{g_{\mathrm{eff}}}{2N}\left[p_{u}^{\dagger}p_{u}^{\dagger}p_{u}p_{u}-p_{l}^{\dagger}p_{l}^{\dagger}p_{l}p_{l}-\left(p_{u}^{\dagger}p_{u}-p_{l}^{\dagger}p_{l}\right)\left(p_{u}^{\dagger}p_{l}+p_{u}p_{l}^{\dagger}\right)+\left(p_{u}^{\dagger}p_{l}-p_{u}p_{l}^{\dagger}\right)\right].\nonumber \\
& &
\label{eq:PH}
\end{eqnarray}
The polaritons $p_{u,l}$ have resonant frequencies of $\Delta_{u,l}$
and experience Kerr nonlinearity with an effective coupling strength
of $\mp\frac{g_{\mathrm{eff}}}{2N}$, which are shown by the first
two terms in the second line. The third and last terms in the second
line describe the mode conversion between the two polariton states.
For a large atom number $N$, a similar treatment can also be extended
to situation where the atoms are not uniformly coupled to the cavity
since the correlation between atoms can be neglected.
\medskip{}

\begin{figure}[htbp]
\centering
\includegraphics[width=.6\linewidth]{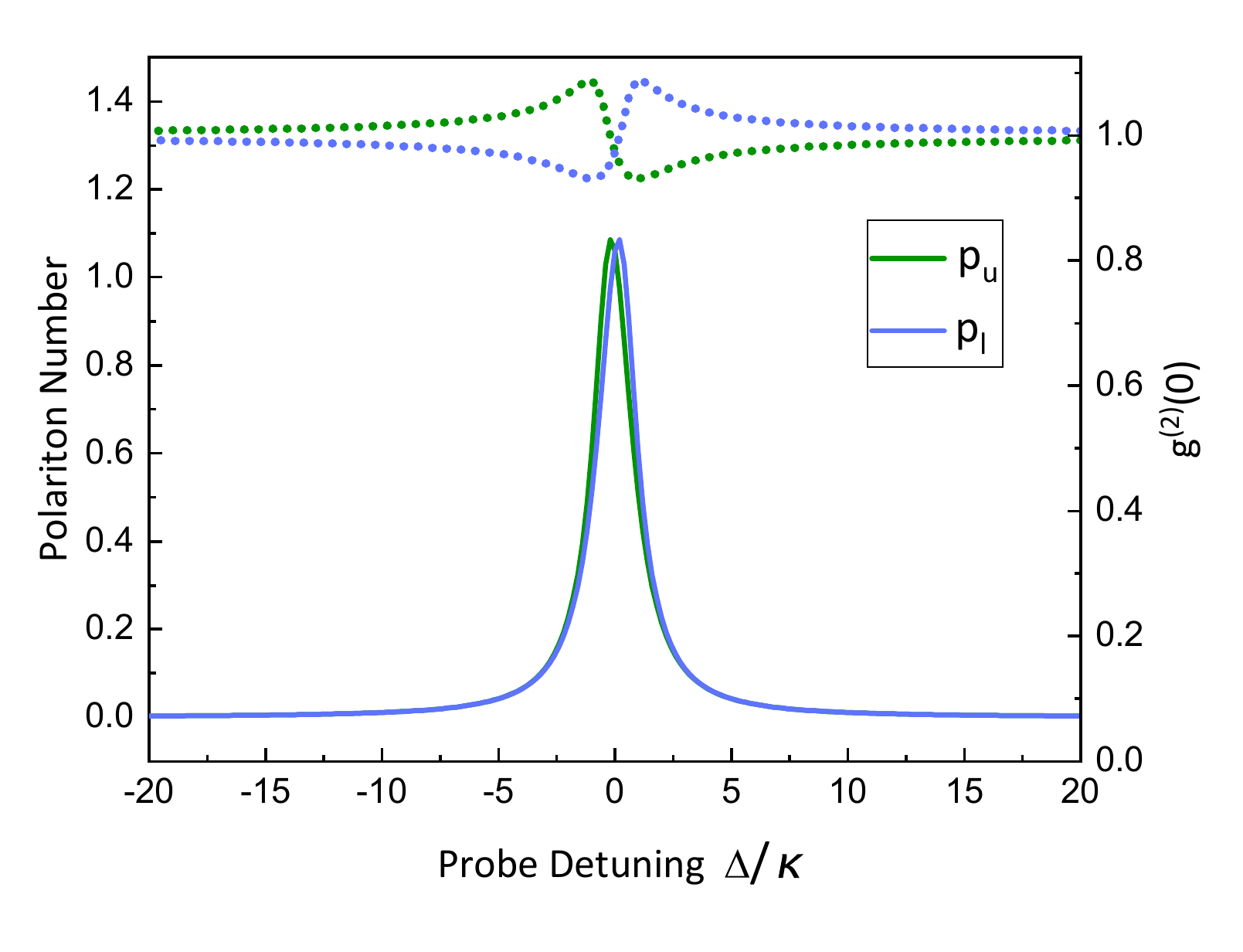}
\caption{Basic properties of a single cavity polariton with self-Kerr nonlinearity. (Solid lines) Relationship between the mean excitation number and the driving
frequency. (Dotted lines) Relationship between the second-order correlation $g^{(2)}\left(0\right)$
and the driving frequency. The calculations are performed with the following parameters:
number of atoms $N=55$, atom-cavity coupling strength $g_{\mathrm{eff}}=8.4947$, driving strength
$\varepsilon_{p}=\sqrt{1.1}$, cavity decay rate $\kappa=1$, and atom
decay rate $\gamma=0.71233$.}
\label{fig:S5}
\end{figure}

\noindent First, we investigate the properties of individual cavity polariton
states under coherent driving. The interaction Hamiltonians are the
self-Kerr (self-phase modulation) terms $H_{I}=-\frac{g_{\mathrm{eff}}}{2N}p_{u}^{\dagger}p_{u}^{\dagger}p_{u}p_{u}$
and $H_{I}=+\frac{g_{\mathrm{eff}}}{2N}p_{l}^{\dagger}p_{l}^{\dagger}p_{l}p_{l}$.
Since $N$ is very large, the nonlinearity of both polariton states
is too weak to induce a significant single photon nonlinear effect, such
as the polariton blockade effect, in neither $p_{u}$ nor $p_{l}$. Figure \ref{fig:S5}
shows the frequency dependence of the intracavity excitation number
and the second-order correlation of each polariton state. The two
curves represent $g_{\mathrm{eff}}=\pm8.4947\:\kappa$,
corresponding to the case of the $p_{l,u}$ mode. Since the Kerr nonlinearities
of different signs shift the polariton resonance in opposite directions,
both curves shift slightly away from zero detuning. Additionally, the
polariton state exhibits anhamonicity, and the energy level of the Fock
state $|n\rangle$ shifts by $\frac{g_{\mathrm{eff}}}{2N}n(n-1)$,
which is demonstrated by the second-order correlation function $g^{(2)}\left(0\right)$
of the field. For the case of $p_{u}$ with negative Kerr nonlinearity
$-\frac{g_{\mathrm{eff}}}{2N}$, the transition frequency between
adjacent Fock states $|n\rangle$ and $|n+1\rangle$ is detuned by
$-\frac{g_{\mathrm{eff}}}{2N}2n$. Therefore, the photon blockade
effect is expected when probed on the blue side, and sub-Poisson distribution
is observed. For the probe on the red side, a Fock state higher
than $|1\rangle$ can be more efficiently excited than the coherent
state; thus, the field shows super-Poisson statistics. A similar analysis
can be applied to the $p_{l}$ polariton, which indicates an opposite
result.
\medskip{}

\noindent Note that the self-Kerr nonlinearity scales inversely
with the atom number $N$. As the number of atoms increases to a sufficiently
large value, the impact of the nonlinearity will become weak so that
the resonance shift and the nonclassical statistics are negligible.
In our calculation, $g_{\mathrm{eff}}=8.4947$ and $N=55$, and only a $5\%$ deviation of $g^{(2)}\left(0\right)$ from $1$ is observed.

\vbox{}

\begin{center}
\textbf{\large S6. Quantum interference between cavity polaritons}\\
\end{center}

\noindent In the strongly coupled cavity-atom system, the hybridization
of the cavity mode and bosonic collective spin mode constitute two
cavity polaritons. They can be excited selectively or simultaneously
by choosing an appropriate frequency of the probe field. Under coherent
field probing on the cavity mode, these two polaritons are simultaneously
probed with coherent fields (Eq.~(\ref{eq:Probe})), described by
the Hamiltonian
\begin{eqnarray}
H_{p} & = & \varepsilon_{p_{u}}\left(p_{u}^{\dagger}+p_{u}\right)+\varepsilon_{l}\left(p_{l}^{\dagger}+p_{l}\right),
\end{eqnarray}
with $\varepsilon_{p_{u,l}}$ being the effective driving strength. In addition to the self-Kerr nonlinearity of individual
polaritons, they also exchange energy with each other (Eq.$\:$\ref{eq:PH}).
Therefore, each polariton is excited in two ways: coherent driving
and coherent conversion from the other polariton. Here, the response
of the polariton under coherent driving is investigated with the experimental
parameters $g_{\mathrm{eff}}=8.4947$ for $\sigma_{-}$ light and $g_{\mathrm{eff}}=4.6424$
for $\sigma_{+}$ light, with the atom number set to $N=55$.
\medskip{}

\begin{figure}[htbp]
\centering
\includegraphics[width=.9\linewidth]{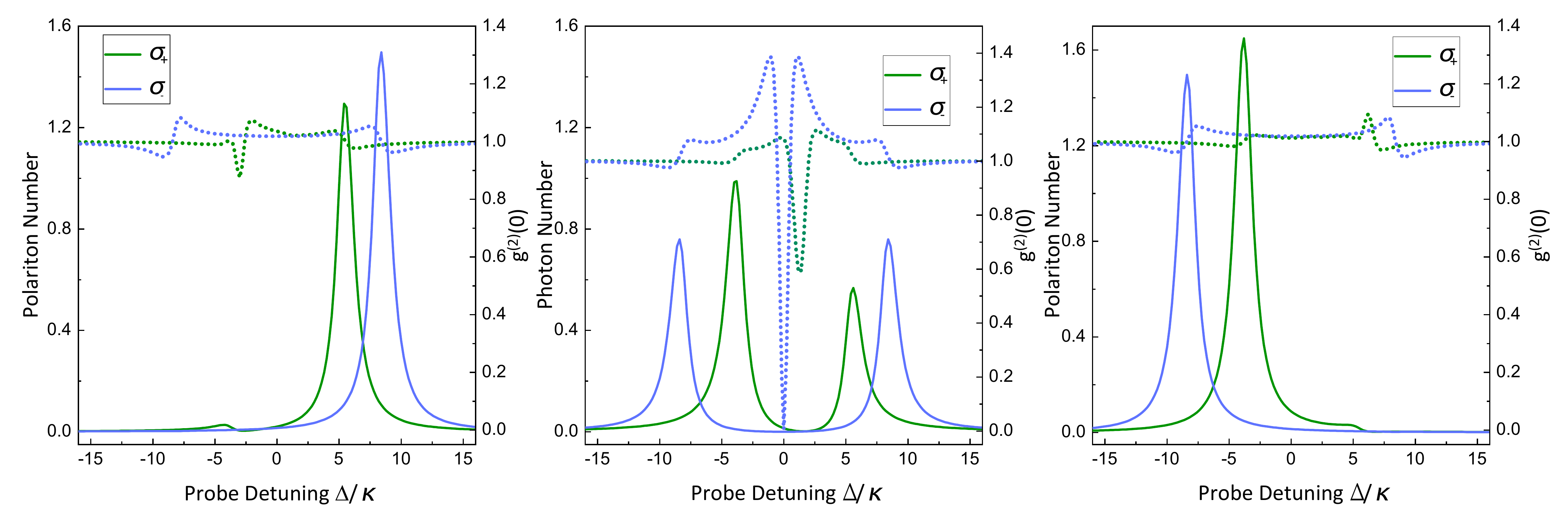}
\caption{Mean excitation number and second-order quantum correlation $g^{(2)}\left(0\right)$
of the cavity mode and cavity-atom modes. Left: polariton state $p_{u}$;
Middle: cavity field; Right: polariton state $p_{l}$. The parameters
are $\kappa=1$, $\gamma=0.7123$, $\varepsilon_{p}=\sqrt{2.2}$, and $N=55$. For the $\sigma_{+}$ case, $g_{\mathrm{eff}}=4.4624$ and
the cavity-atom detuning $\Delta_{ac}=4.6438$. For the $\sigma_{-}$
case, $g_{\mathrm{eff}}=8.4947$ and the atom is in resonance with
the cavity. Solid line: mean excitation number (left axis). Dashed line: second-order
correlation function $g^{(2)}\left(0\right)$ (right axis)}
\label{fig:S6}
\end{figure}

\noindent Since the probe field to the cavity can excite two polariton states
simultaneously, an interesting quantum effect arises due to the interference
between the optical emissions of $p_{u,l}$. As shown in Figure \ref{fig:S6},
the frequency dependence of the polariton excitation and $g^{(2)}\left(0\right)$
are investigated numerically. Since the frequencies of the polariton states
are different, when the probe is near resonant with one polariton
state, the influence of the other polariton state is negligible.
As shown in Figure \ref{fig:S6} (middle and right panels), the spectrum shows similar
behaviors to Figure 1 in the main text when the probe is near resonant with $p_{u}$
or $p_{l}$. In particular, $g^{(2)}\left(0\right)$ shows a small
modification due to the self-Kerr effect of polaritons. It should
also be noted that when the probe field is tuned to resonate with the
other polariton, for example, $p_{l}$ in the left panel of Figure \ref{fig:S6},
a significant $g^{(2)}\left(0\right)$ pattern is also observed for
the $p_{u}$ polariton even though it is not efficiently excited. More
interestingly, the $g^{(2)}\left(0\right)$ pattern of $p_{u}$ is
more pronounced than that of the resonant excitation. Such an unusual
phenomenon could be accounted for by the interference between the directly
excited $p_{u}$ and the conversion from $p_{l}$ since $p_{l}$
is efficiently excited. These two processes both possess Kerr nonlinearity
and alter the phase of Fock states higher than $|1\rangle$ in different
manners, thus suppressing or enhancing the components of high-order
Fock states according to the driving frequency, leading to bunching
or anti-bunching of the polariton field.
\medskip{}

\noindent In experiments, the quantum statistics can only be measured for the
cavity output field, which is a combination of both polariton states. The middle panel of Figure \ref{fig:S6} shows the equivalent cavity photon number
and $g^{(2)}\left(0\right)$ against the probe frequency. The two
peaks of each curve clearly show the resonances of polariton states.
As an example, we explain the $\sigma_{-}$ case in which the cavity
and atoms share the same frequency. As shown by the blue curve in
Figure \ref{fig:S5}, the curve can be divided into three regions.
On the left, the probe field is near-resonance with $p_{l}$ and far
off-resonance with $p_{u}$. The cavity field is mainly contributed
by $p_{l}$; thus, its second-order correlation behaves the same with
$p_{l}$ (shown by the right panel of Figure \ref{fig:S6}). Similarly, the curve on
the right side is the same as $p_{u}$ in the left panel of Figure \ref{fig:S6}.
For the case of a driving frequency near the center of the polariton
frequencies (the middle of the middle panel of Figure \ref{fig:S6}), $p_{u}$ and $p_{l}$
are simultaneously excited with approximately equal strength, and the
output cavity field is now a superposition of the emission from two
polariton states. In this case, significant interference is expected.
Due to the opposite sign of $p_{u}^{\dagger}p_{u}^{\dagger}p_{u}p_{u}$
and $p_{l}^{\dagger}p_{l}^{\dagger}p_{l}p_{l}$, the most significant
term in the Fock space, i.e., the two-photon state $|2\rangle$ have opposite
phases. Therefore, the component of $|2\rangle$ is strongly suppressed
after the projection of the polaritons to the cavity field if the
polaritons are symmetrically excited. This phenomenon is similar to
the unconventional photon blockade~~\cite{Flayac:2017}, by which significant
anti-bunching can be observed with only very weak nonlinearity. The
most significant destructive interference is expected when excitations
of two polariton states are the same. In our calculation, a nearly
zero $g^{(2)}\left(0\right)$ is observed with nonlinearity as low
as $g_{\mathrm{eff}}/N=0.077\:\kappa$, as shown by the dotted blue curve.
If the probe field is shifted slightly away from the maximum point, the
nonlinear phases induced by two polaritons change, leading to constructive interference and bunching of the cavity field with $g^{(2)}\left(0\right)>1$.
By exploiting the interference between cavity polaritons, the analysis
clearly reveals the mechanics of quantum statistics in cavity QED system
containing many emitters without numerically calculating the system
in an exponentially scaled Hilbert space.

\vbox{}

\begin{center}
\textbf{\large S7. Nonreciprocity of the cavity polaritons}\\
\end{center}

\noindent In the following, we compare the two cases where the atoms are initially
prepared in different Zeeman sublevels and probed by
light with different circular polarizations, and show the nonreciprocity
of the cavity-polaritons. As presented in the main text, the coupling
strength $g$ for the $\sigma_{\pm}$-polarized cavity mode is different
when the atoms are prepared to a Zeeman state that breaks the time-reversal
symmetry. According to the experimental parameters, the numerical
results in Figure \ref{fig:S6} show non-reciprocal mean-field and second-order
correlation functions, as the spectra are different for the $\left(+z,\sigma_{+}\right)$
and $\left(-z,\sigma_{-}\right)$ probes.
\medskip{}

\noindent By probing the system from different directions with different polarizations,
the non-reciprocal excitation of the polaritons is expected, as can
be measured by the non-reciprocal transmission of the cavity field.
Such non-reciprocal phenomena mainly result from the difference
in the atom-photon coupling strength or the cooperativity of the system.
Additionally, compared with the $\sigma_{-}$ case, the transition
between atom energy levels that couples with the $\sigma_{+}$ light
is shifted away from the cavity resonance due to the stark shift $\Delta_\text{Stark}$,
which agrees well with the experimental results. Therefore, the resonances
of the polaritons are asymmetric for $\sigma_{+}$ coupling.
\medskip{}

\noindent In addition, the quantum statistics of the polaritons are also non-reciprocal.
Since they have different resonant frequencies, nonclassical statistics
appear only when they are nearly resonant-excited, while they almost behave
classically for far-off resonant excitation. Such non-reciprocal quantum
statistics of polaritons are shown by the blue ($\sigma_{-}$) and
green ($\sigma_{+}$) dashed curves. For example, if the system is
probed by light with frequency detuning $\Delta/\kappa=7$, the polariton
$p_{u}$ of $\sigma_{-}$ coupling shows sub-Poisson statistics, while
the $\sigma_{+}$ coupling shows super-Poisson statistics (left panel of
Figure \ref{fig:S6}). The non-reciprocal quantum statistics of polaritons
are the most significant when the probe is near the resonances.

\vbox{}

\begin{center}
\textbf{\large  S8. Non-reciprocal quantum interference}\\
\end{center}

\noindent In the last section, we focused on the quantum interference between
cavity polaritons, which is reflected by the second-order correlation
$g^{(2)}\left(0\right)$ of the cavity field. It was noted that
the quantum interference of two paths requires them to have nearly
the same amplitude. In our system, the most significant parameter
that influences the excitation of two cavity polaritons is the relative
detuning of the probe field. When probed with frequency near the
center of the two polaritons, strong quantum interference is observed,
indicated by the significant interference pattern of the $g^{(2)}\left(0\right)$
of the cavity field (dashed curve, Figure \ref{fig:S6}). Due to the
AC-Stark shift, the resonances of the $\sigma_{-}$ polaritons are
symmetrically spaced on different sites of the cavity resonance, while
the resonances of the $\sigma_{+}$ polaritons are asymmetric with
$p_{l}$ near the cavity resonance and $p_{u}$ far from it. Therefore,
the strongest quantum interference occurs at $\Delta=0$ for $\sigma_{-}$,
where the two polaritons are balanced excited. For $\sigma_{+}$,
the probe frequency should be shifted toward $p_{u}$ to balance
the excitations. At the balanced point, the destructive interference
greatly suppresses the population of large photon-number states, leading
to anti-bunching of the cavity output field. With a small detuning,
the constructive interference results in bunching of the cavity
field.

\begin{figure}[htbp]
\centering
\includegraphics[width=.8\linewidth]{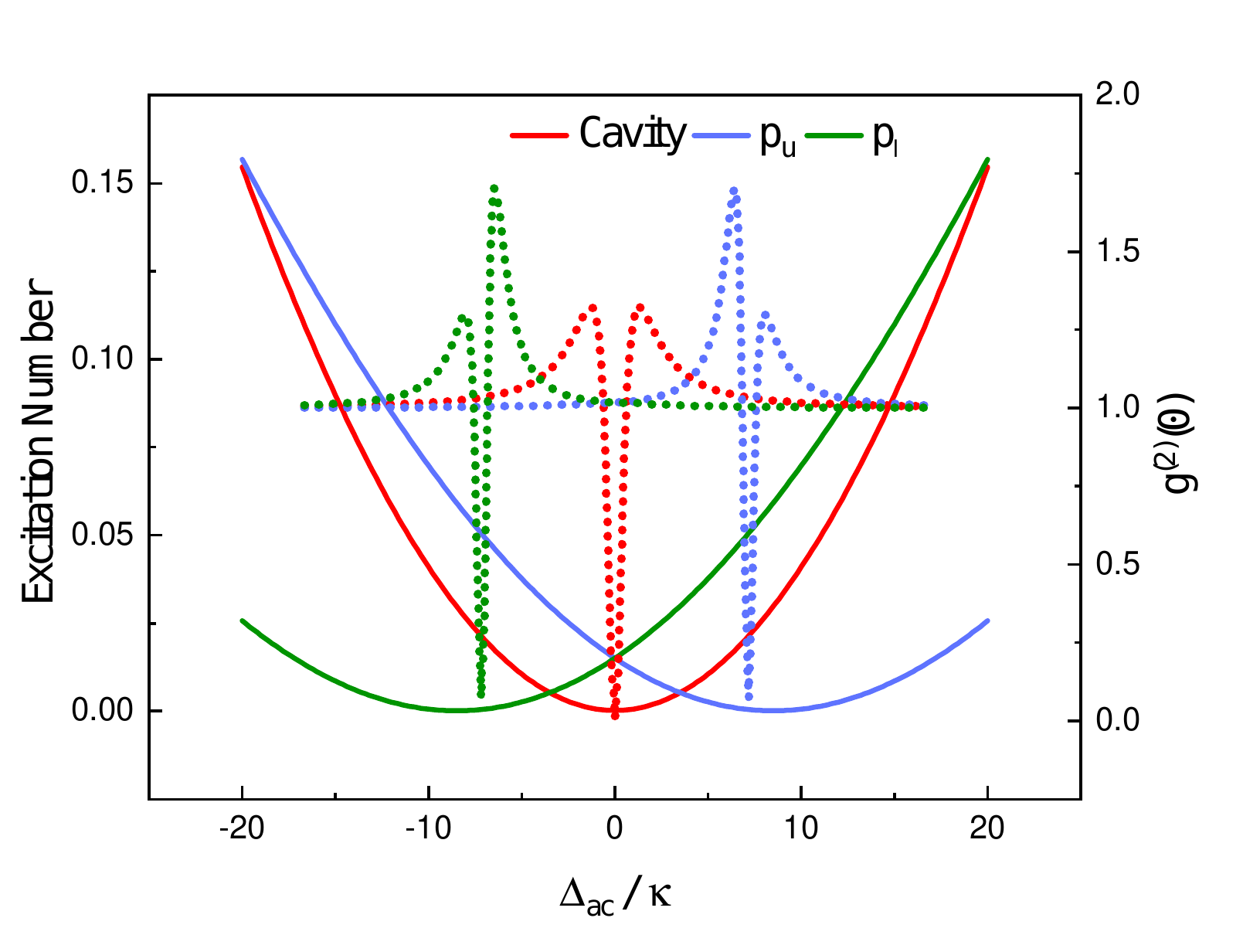}
\caption{The mean excitation number and the second-order correlation function
$g^{(2)}\left(0\right)$ of the polaritons versus the cavity-atom
detuning $\Delta_{ac}$. The parameters are $\kappa=1$, $\gamma=0.7123$,
$\varepsilon_{p}=\sqrt{2.2}$, $N=55$, and $g_{\mathrm{eff}}=8.4947$.}
\label{fig:S7}
\end{figure}

\medskip{}

\noindent The nonreciprocity of the quantum interference depends on the cavity-atom
detuning $\Delta_{ac}$, which is shown in Figure \ref{fig:S7}. Under
resonant probe $\Delta=0$, $p_{u}$ is efficiently excited for negative
detunings and suppressed for positive detunings (blue curve, Figure \ref{fig:S7}),
vice versa for $p_{l}$. Its second-order correlation function $g^{(2)}$$\left(0\right)$
shows a significant interference pattern at an appropriate $\Delta_{ac}>0$,
at which $p_{u}$ is suppressed and $p_{l}$ is excited. Due to the
coherent conversion from $p_{l}$ to $p_{u}$, these two components
are balanced at a critical value of $\Delta_{ac}$ and interfere with each other,
leading to strong bunching and anti-bunching of the polariton. The
solid and dashed red curves show the mean photon number and $g^{(2)}\left(0\right)$
of the cavity field, respectively, which also shows strong quantum
interference.

\begin{figure}[htbp]
\centering
\includegraphics[width=.8\linewidth]{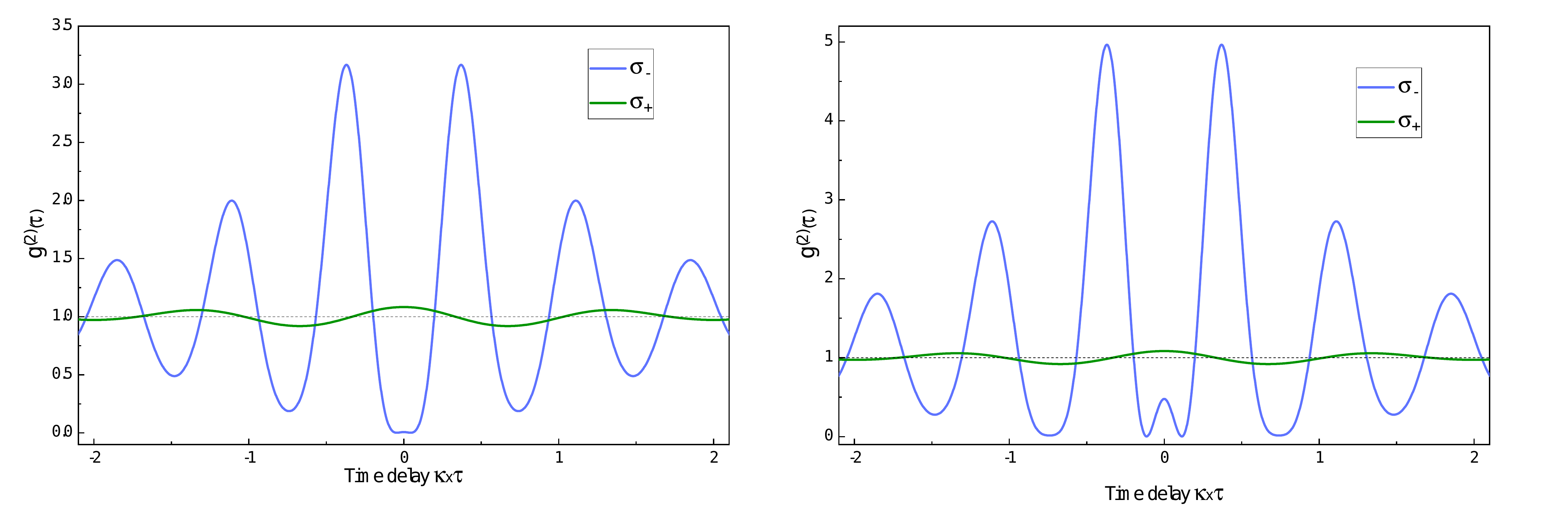}
\caption{Second-order correlation function $g^{(2)}\left(\tau\right)$ of the
cavity field for $\sigma_{+}$ and $\sigma_{-}$ driving. Left: $N=55$. Right: $N=35$.}
\label{fig:S8}
\end{figure}
\medskip{}

\noindent In our experiment, the nonreciprocity of the quantum interference
is tested at $\Delta=0$. For resonant probe
$\Delta=0$, strong interference is indicated by $g^{(2)}\left(0\right)$,
with $g^{(2)}\left(0\right)>1$ for $\sigma_{+}$ and $g^{(2)}\left(0\right)<1$
for $\sigma_{-}$ (Figure \ref{fig:S8}). In this case, $g^{(2)}\left(\tau\right)$
oscillates with the time delay $\tau$, due to the beating between
the polariton state. The oscillation period is proportional to the
frequency difference of the two polariton states.

\vbox{}

\begin{center}
\textbf{\large  S9. Discussion}\\
\end{center}

\begin{figure}[htbp]
\centering
\includegraphics[width=.6\linewidth]{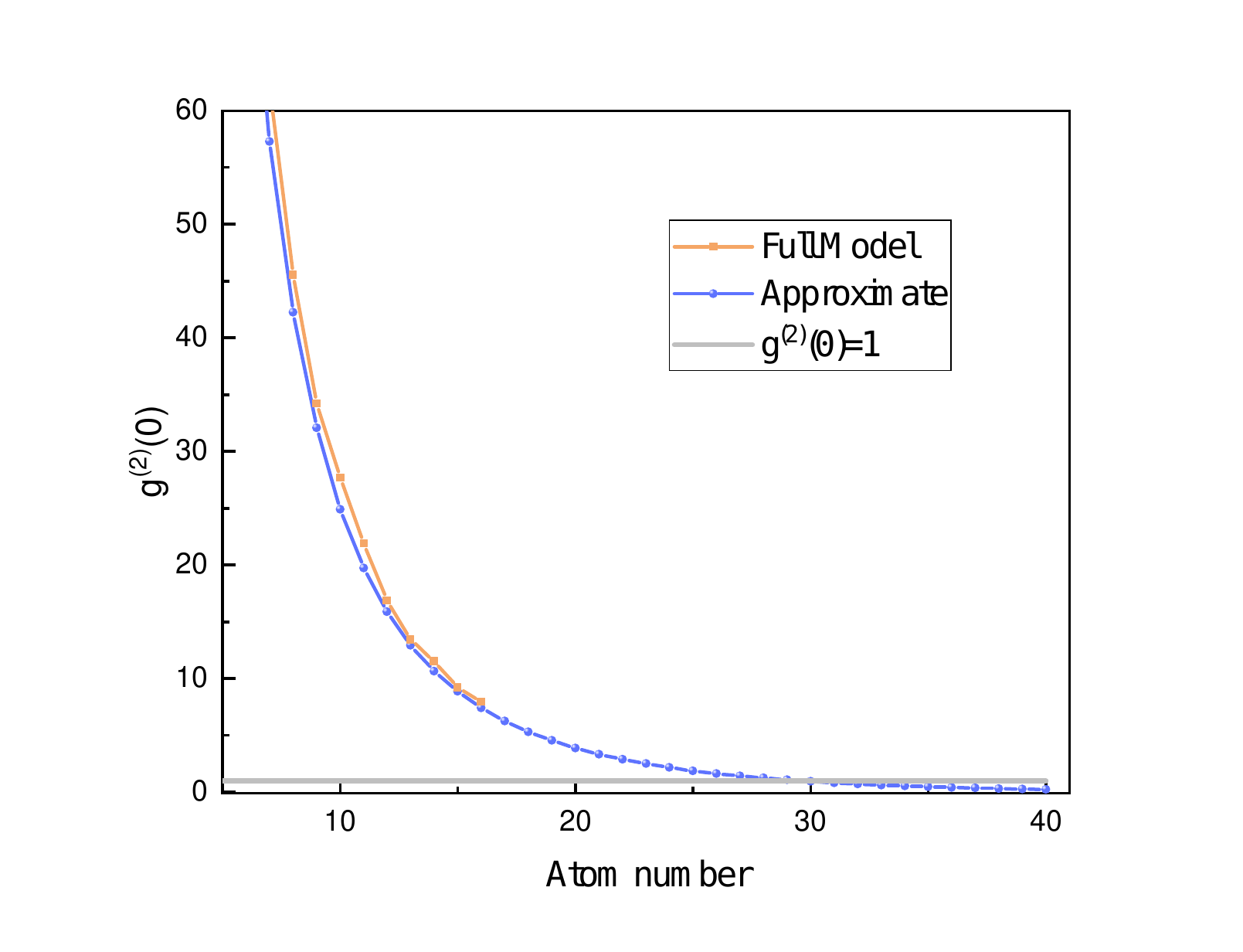}
\caption{Comparison between the numerical results of the full model and the
approximate model. In the calculations, the cooperativity $C$ is
kept the same by increasing the atom number $N$ and reducing the
coupling strength $g$. The parameters used are $g\sqrt{N}=8.4316$,
$\kappa=1$, $\gamma=0.7123$, $\varepsilon_{p}=\sqrt{2.2}$, and $\Delta=\Delta_{ac}=0$.
Blue line: results by the model in Eq.$\:$(\ref{eq:PH}). Orange
line: results by the model in Eq.$\:$(\ref{eq:cqed}). As the atom
number $N$ increases, the results of the approximate mode based on
the bosonic treatment of the collective atomic spin approach the
results of the full Hamiltonian.}
\label{fig:S9}
\end{figure}

\noindent To summarize, the non-reciprocal transmission mainly depends on the
cooperativity of the system for the $\sigma_{\pm}$ mode that can
only be separately probed from different directions, while the non-reciprocal
quantum statistics are very sensitive to the atom-cavity detuning due
to quantum interference between polariton states. Compared with single-emitter
cavity QED, the multiatom system can achieve larger cooperativity
via the collective effect, which is beneficial for applications, such
as high-isolation ratio non-reciprocal devices with higher saturation
power. For the study of fundamental physics, the quantum property
of cavity polaritons behaves differently, which depends on the atom
number $N$, coupling strength $g$ and driving field $\varepsilon_{p}$.
A high cooperativity system can be constructed either by a few atoms
coupled to a cavity with strong interaction strength or many atoms
with weak interaction strength. These two systems have distinct quantum statistics,
as seen in Figure \ref{fig:S9}. The blue line in Figure \ref{fig:S9}
shows the second-order correlation function of the cavity field in
systems with fixed $g_{\mathrm{eff}}$. As the number of atoms increases,
the statistics of the cavity field change from super-Poisson
to sub-Poisson. For a very large atom number $N$, the $g^{(2)}\left(0\right)$
of the cavity field increases and approaches $1$ due to the decrease
in the nonlinearity $\frac{g_{\mathrm{eff}}}{2N}$ of a single cavity
polariton. Our experimental study lies in the moderate region where
the cavity polaritons are not saturated and their nonlinearity is
relatively large. It should be noted that the treatment that includes
only the first two orders of Eq.~(\ref{eq:Taylor}) should be extended
to higher-order series under strong excitation, offering a new platform
to study high-order nonlinear effects.

\vbox{}

\begin{center}
\textbf{\large S10. Comparison with full-numerically result}\\
\end{center}

\noindent In the above analysis, the system with effective total atomic spin
$N/2$ is approximately treated as a bosonic mode. At the low-excitation
limit, atomic nonlinearity in the few-atom ensemble is approximated
to the first order of $1/N$. To verify the validity of the low-excitation
approximation, we compare the results of the above model and the numerical
solution of the full Hamiltonian {[}Eq.~(\ref{eq:cqed}){]}. Because the
dimension of the Hilbert space of the system exponentially increases
with the number of atoms ($N$) as $d=M\times2^{N}$, where $M$ is
the truncated dimension of the cavity mode, the computation complexity
increases rapidly with increased pump power and number of atoms, and the
numerical simulation is limited to a small $N$. Here, we use the
quantum trajectory approach to simulate the system evolution~\cite{Johansson:2013},
which is less demanding for computation resources since only the
pure state (dimension $d$) is considered in the quantum trajectory
simulation instead of the density matrix (dimension $d^{2}$). We fix the value of $g_{\mathrm{eff}}=g\sqrt{N}$, thus fixing the cooperativity
$C$, and change the atom number $N$. In Figure \ref{fig:S8}, we plot
the second-order correlation function $g^{(2)}\left(0\right)$ of
the cavity output calculated from both the approximate and full models.
The difference between the two curves decreases as the number of atoms increases,
which verifies the validity of our analysis based on the approximate
model.

\vbox{}

\begin{center}
\textbf{\large S11. Simulation of reconfiguration of optical isolator}\\
\end{center}

\begin{figure}[htbp]
\centering
\includegraphics[width=.6\linewidth]{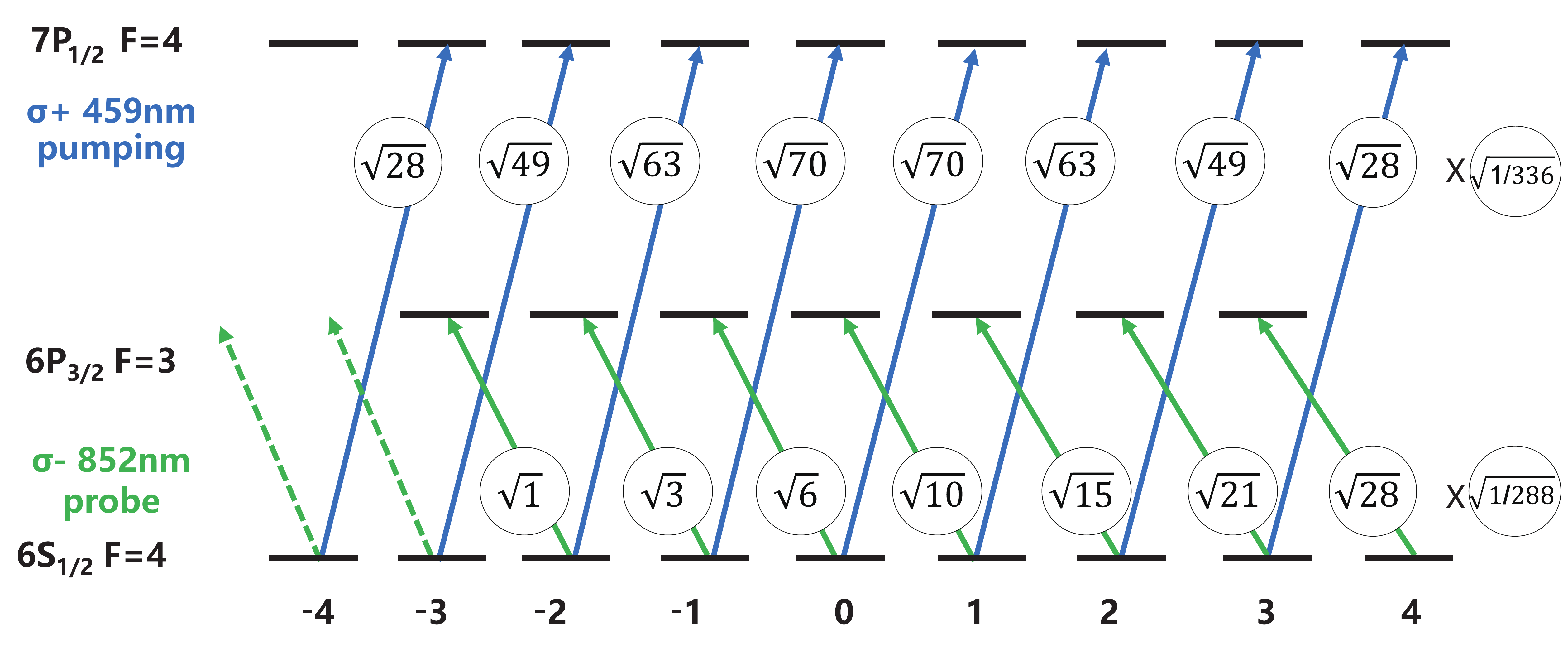}
\caption{Energy levels and optical pumping procedure. Atoms initially populated at $|6S_{1/2},F=4,mF=-4\rangle$. The probe field
resonates to transition between $|6S_{1/2},F=4\rangle$ and $|6P_{3/2},F=3\rangle$, and the internal state of the atom is controlled
by the 459-nm pump laser which couples to the transition between $|6S_{1/2},F=4\rangle$ and $|7P_{1/2},F=4\rangle$. The numbers inside the circles are the corresponding reduced dipole matrix elements.}
\label{fig:S10}
\end{figure}

\noindent In our experiment, the switch of the isolation direction is realized by controlling
the populations of the internal state of the atoms. As shown in Figure \ref{fig:S10}, the probe field (green arrows) 
is resonant to the $|6S_{1/2},F=4\rangle\leftrightarrow|6P_{3/2},F=3\rangle$ transition. The populations of the internal state of the atom are controlled
by the 459-nm pump laser (blue arrows), which is resonant to the transition $|6S_{1/2},F=4\rangle\leftrightarrow|7P_{1/2},F=4\rangle$. The probe field with $\sigma_{-}$ polarization transmits through the
cavity when the atom is populated at $m_{F}=-4\ \text{or}\ -3$, while it is blockaded
when the atom is populated in $m_{F}=4$ due to vacuum Rabi splitting. It should be noted that the switch between transmission and blockade is asymmetric depending on the
initial state of the atom. In our experiment, we use an ensemble
of atoms that strongly couple to the atomic transition. For the
transmission case, all the atoms are required to be in the $m_{F}=-4$
and $m_{F}=-3$ states. However, to block the probe field, only a
small portion of atoms on $m_{F}\geq-2$ can induce significant
Rabi splitting. If we start with the transmission case where all
atoms are prepared at $m_{F}=-4$, the internal state of the atoms
will be gradually pumped to $m_{F}\geq-2$ under a $\sigma_{+}$ pump field. 
The transmission of the probe will be switched off
quickly as long as the states of several atoms are transferred.
The larger the atom number and the stronger the vacuum coupling strength, the faster
the switch-off speed. In contrast, when we start from the blockade
state with all atoms prepared at $m_{F}=4$, transferring 
all the atoms to $m_{F}=-3$ and $m_{F}-4$ with a $\sigma_{-}$ pump laser
will take a sufficiently long time. The smaller
the atom number and the weaker the vacuum coupling strength, the slower the switch-on
speed. 

\begin{figure}[htbp]
\centering
\includegraphics[width=.8\linewidth]{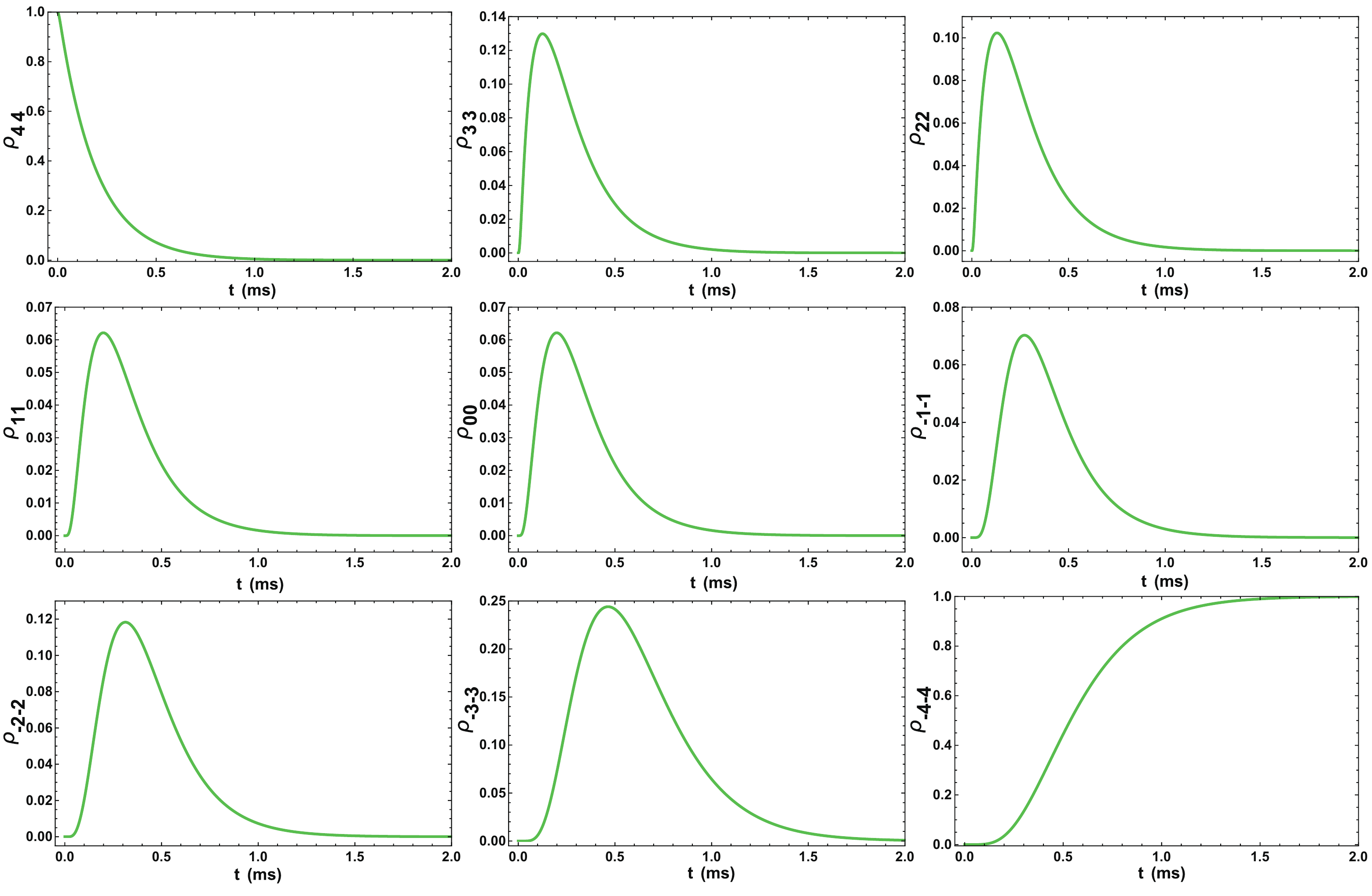}
\caption{Atomic population ($\rho_{ii}$) versus time during the turn-on procedure. A $\sigma_{-}$ 459-nm pump laser is used to continuously pump the atoms from $m_{F}=4$ to $m_{F}=-4$ with an optical pumping rate equal to 2.3$\Gamma$. The 459-nm pumping light with polarization $\sigma_+$  is turned on at $t=0$.}
\label{fig:S11}
\end{figure}
\medskip{}

\begin{figure}[htbp]
\centering
\includegraphics[width=.8\linewidth]{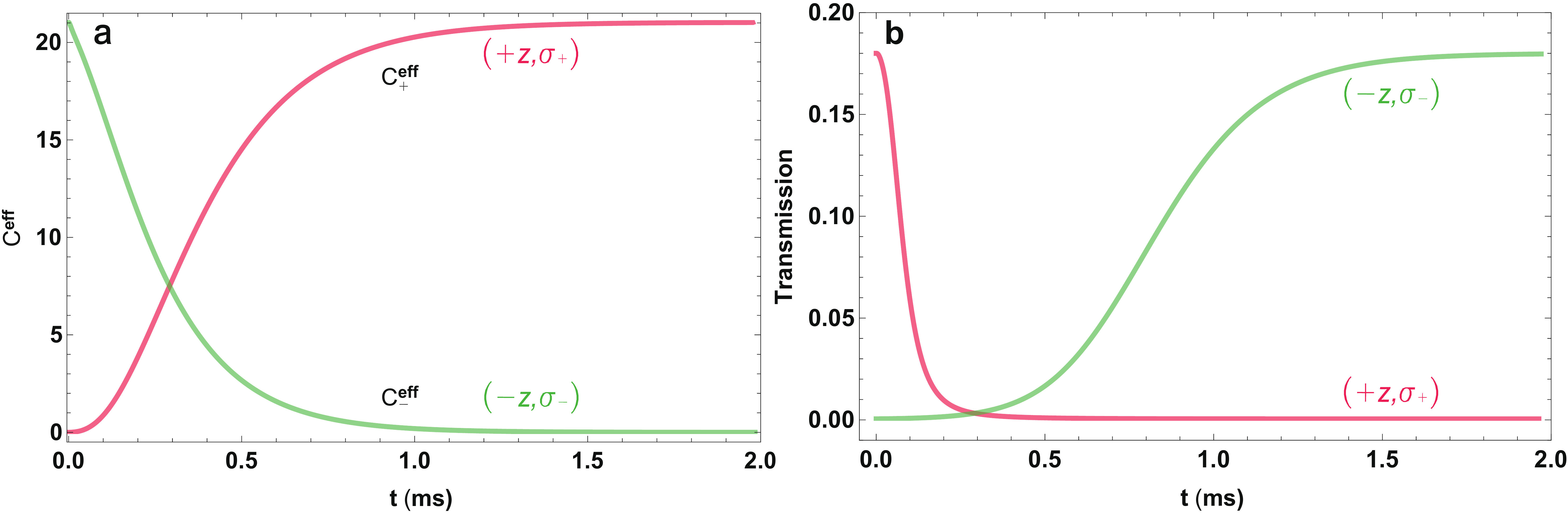}
\caption{The effective cooperativity and transmission versus time during the switch procedure. A 459-nm pump laser is used to continuously pump the atoms between $m_{F}=-4$ and $m_{F}=4$ with the same optical pumping rate in Figure 4. (\textbf{a}) The variance of effective cooperativity as the shut-off (red curve) and turn-on (green curve) procedures. (\textbf{b}) The corresponding transmission during the same procedures as in (\textbf{a}). The 459-nm pumping light with polarization $\sigma_+$ is turned on at $t=0$.}
\label{fig:S12}
\end{figure}

\noindent To quantitatively study the switching process, we theoretically simulated the optical
pumping process by taking all the experimental parameters. After switching the
polarization of the 459-nm pump light, the variance of the atomic population on each Zeeman
state can be obtained by using the rate equations, and the results are shown in Figure \ref{fig:S11}.  
A maximum Rabi frequency of 2.3$\Gamma$ for the 459-nm optical light is used, 
and $\Gamma$ is the decay rate of $|7P_{1/2},F=4\rangle$.
In this simulation, we only consider the effect of the pump laser and neglect that of the probe light since it is sufficiently
weak. The variance of the effective parameter of cooperativity versus time is then obtained by
\begin{eqnarray}
C_{\pm}^{\mathrm{eff}} & = & N\sum_{i}\rho_{ii}g_{i,\pm}^{2}/2\kappa \gamma,
\end{eqnarray}
where $\rho_{ii}$ is the population of $m_{F}=i$, $N$ is the total
atom number, and $g_{i,\pm}$ is the coupling strength between $m_{F}=i$
and $m_{F'}=i\pm1$. The dependence of $C_{\pm}^{\mathrm{eff}}$ on time
is shown in Figure \ref{fig:S12}a. The transmittance of the probe field can be calculated by 
using Eq. (1) in the main text, and the results are given in Figure \ref{fig:S12}b.
The theoretical results are in good agreement with the experimental results.


\end{document}